\newif\ifanonymous
\anonymousfalse

\documentclass[11pt,letterpaper]{article}

\usepackage{fullpage}
\usepackage[utf8]{inputenc}
\usepackage[T1]{fontenc}
\usepackage[UKenglish]{babel}
\usepackage{lmodern}
\usepackage{microtype}

 \usepackage{array}
 \usepackage{multirow}

\usepackage{amsmath,amssymb,amsthm}
\usepackage{booktabs}
\usepackage{algorithmic}
\usepackage{pifont}
\usepackage{tikz}
\usetikzlibrary{fit}

\usepackage[colorlinks=true,allcolors=blue]{hyperref}
\usepackage[nameinlink,noabbrev]{cleveref}

\usepackage{amsfonts, amsmath, amsthm, amssymb}
\usepackage{bm}
\usepackage[utf8]{inputenc}
\usepackage{hyperref}
\usepackage{xcolor}
\usepackage{xspace}
\usepackage{url}
\usepackage{cleveref}
\usepackage{mdframed}
\usepackage{tikz}
\usetikzlibrary{positioning, arrows.meta, shapes.geometric, backgrounds}

\usepackage{url}
\usepackage{xcolor}
\usepackage{soul}
\usepackage[colorinlistoftodos]{todonotes}
\usepackage{tcolorbox}
\usepackage{graphicx} 
\usepackage[ruled,vlined,linesnumbered]{algorithm2e}
\usepackage{rotating} 
\usepackage[dvipsnames]{xcolor}
\usepackage{enumitem}
\usepackage{cite}
\newcommand{\pparagraph}[1]{\medskip \noindent \textbf{#1.}}
\hypersetup{colorlinks,linkcolor={black},citecolor={teal},urlcolor={}}

\usepackage{tcolorbox}
 \usepackage[export]{adjustbox}
 \usepackage{varwidth}
 \usepackage{lipsum}
 \usepackage{mdframed}
\usepackage{subcaption}

\ifdefined\publish
  \newcommand{\missing}[1]{}
  \newcommand{\alireza}[1]{}
  \newcommand{\zhipeng}[1]{}
    \newcommand{\istvan}[1]{}
\else
  \newcommand{\missing}[1]{\textcolor{orange}{#1}}
  \newcommand{\alireza}[1]{\textcolor{blue}{\textbf{Alireza:} #1}}
  \newcommand{\zhipeng}[1]{\textcolor{red}{\textbf{Zhipeng:} #1}}
  \newcommand{\istvan}[1]{\textcolor{orange}{\textbf{Istvan:} #1}}
\fi

\newcommand{\sys}{\textsf{Proof of Source of Funds}\xspace} 

\newcommand{\syss}{\textsf{PoSoF}\xspace}

\newcommand{\eg}{\textit{e.g.,} }
\newcommand{\ie}{\textit{i.e.,} }

\newcommand{\NN}{\mathbb{N}} 

\newcommand{\rel}{\mathcal{R}}
\newcommand{\adv}{\mathcal{A}}

\newcommand{\crs}{\mathsf{crs}}
\newcommand{\td}{\mathsf{td}}

\newcommand{\prover}{\mathcal{P}}
\newcommand{\verifier}{\mathcal{V}}
\newcommand{\ext}{\mathcal{E}} 
\newcommand{\setup}{\mathsf{Setup}}
\newcommand{\prove}{\mathsf{Prove}}

\newcommand{\verify}{\mathsf{Verify}}

\newcommand{\simulate}{\mathsf{Simulate}}
\newcommand{\total}{\mathsf{total}}

\newcommand{\negl}{\mathsf{negl}}

\newcommand{\AS}{\mathsf{AS}}

\usetikzlibrary{positioning, arrows.meta}

\theoremstyle{definition}
\newtheorem{definition}{Definition}

\newcommand{\tx}{\mathsf{tx}}
\newcommand{\sk}{\mathsf{sk}}
\newcommand{\pk}{\mathsf{pk}}

\newcommand{\ori}{\mathsf{PoSoF}}

\newtheorem{remark}{Remark}

\newcommand{\addr}{\mathsf{addr}\xspace}

\usepackage[dvipsnames]{xcolor}

\hypersetup{
  colorlinks=true,
  urlcolor=black,
  linkcolor= Blue,
  citecolor=OliveGreen
}

\title{\bf Proof of Source of Funds \\ Efficient On-chain Provenance of Cryptoassets}

\author{
Alireza Kavousi\\
University College London\\
\texttt{a.kavousi@cs.ucl.ac.uk}
\and
István András Seres\\
Eötvös Loránd University\\
\texttt{seresistvanandras@gmail.com}
\and
Zhipeng Wang\\
The University of Manchester\\
\texttt{zhipeng.wang@manchester.ac.uk}
}

\date{ }

\begin{document}
\maketitle

\begin{abstract}
Regulatory compliance is increasingly mandatory for decentralized finance and privacy-enhancing technologies. However, current compliance methodologies rely either on fragile binary inclusion/exclusion lists or continuous, retroactive graph analysis by centralized blockchain intelligence firms. This traditional approach strips honest users of their financial privacy, leads to false positives and negatives, and forces decentralized platforms to bear the heavy burden of on-chain transaction monitoring. In this work, we propose a paradigm shift: moving from platform-side surveillance to \textit{user-side provenance}. We introduce \sys\ (\syss), a novel cryptographic framework that shifts the computational and analytical burden to the individual users. Rather than the platform tracing funds, the user locally generates a zero-knowledge proof demonstrating that their deposit originates exclusively from a set of compliant sources. The platform is thus relieved of chain-analysis duties, requiring only a constant-time, $O(1)$ verification to enforce admission control.

To achieve this, we formulate a unified temporal Directed Acyclic Graph (DAG) abstraction that formalizes both UTXO and account-based ledger histories within a generalized value-flow model. Users extract a compliant sub-DAG of their transaction history and utilize Incrementally Verifiable Computation (IVC) to prove rigorous state-transition predicates that protect against various attack vectors. Crucially, \syss\ provides verifiable cryptographic provenance; it guarantees the legitimacy of the funds without leaking the intermediate transaction topology, intermediary addresses, or even the specific origins utilized. We formally define the security properties of \syss\ and evaluate an Ethereum-compatible prototype. Our benchmarks demonstrate that fully private, proactive compliance is highly practical, requiring only $\sim 1.8$\,s to incrementally update a user's \syss\ proof per new transaction, and a constant-time $\sim 1.5$\,ms ($\sim 800$k gas) for final on-chain EVM verification.
\end{abstract}

\section{Introduction}

\label{sec:intro}

In traditional finance, user privacy and regulatory compliance are managed by trusted intermediaries, such as banks, which act as centralized safeguards. Decentralized Finance (DeFi) eliminates these intermediaries, enabling permissionless, peer-to-peer transactions. However, this architectural shift fundamentally challenges traditional regulatory frameworks. As public blockchains inherently default to transparent pseudonymity—where identities are masked but all transaction histories are fully visible—users are effectively forced to rely on cryptographic privacy pools (\eg Zcash~\cite{sasson2014zerocash}, Tornado Cash~\cite{pertsev2019tornado}, Railgun~\cite{railgun-github}, etc.) to protect their financial data from public surveillance.

While these privacy pools successfully restore user anonymity, they break the ecosystem's prevailing compliance paradigm. Currently, decentralized protocols and centralized exchanges rely heavily on continuous, \emph{retroactive graph analysis} conducted by third-party blockchain intelligence firms (\eg Chainalysis, TRM Labs) to trace transaction histories and flag suspicious interactions~\cite{chainup_kyt_2026}. This reactive approach is fundamentally limited; it is exponentially expensive~\cite{wu2023tracer}, suffers from high false-positive rates, and crucially, fails to prevent illicit funds from entering a protocol in the first place. Furthermore, because privacy pools obfuscate the on-chain transaction graph, they have been increasingly exploited for money laundering and sanction evasion, culminating in severe regulatory backlash and the US Treasury’s sanctioning of Tornado Cash in 2022~\cite{ofac-tornado-cash-2022}. Recent empirical work
demonstrates that adversaries can systematically route illicit funds
through fresh, unflagged addresses to evade both blocklists and
taint-propagation monitors~\cite{liu2026evasion}.

Recent academic and industry efforts to design compliant privacy protocols~\cite{buterin2024blockchain, baranski2023haze, beal2024derecho} have attempted to solve this by introducing zero-knowledge \emph{membership sets} (binary inclusion or exclusion proofs). In these models, users prove their deposits either belong to a curated allowlist or do not originate from a publicly known blocklist. However, both architectures offer limited protection. Adversaries easily evade them by transferring funds to freshly generated, unflagged addresses prior to deposit. Furthermore, these retroactive designs open the door to \emph{tainting attacks}~\cite{wang2023blockchain}, where malicious actors intentionally send illicit dust to honest users' addresses. By forcibly intertwining illicit fund with the legitimate transaction graph, adversaries break both inclusion and exclusion proofs, creating a severe regulatory dilemma: monitors must either flag all tainted accounts, unjustifiably freezing honest users' funds~\cite{sahu2023sede}—or weaken their compliance filters, allowing illicit funds to easily bypass the system. Finally, the dynamic nature of these lists also demands real-time synchronization, which is highly problematic and expensive for on-chain verifiers~\cite{beal2024derecho}.


Given the inherent failures of retroactive surveillance and binary exclusion-list architectures, we propose a paradigm shift. In this work, we introduce \sys (\syss), a cryptographic framework that leverages \emph{on-chain fund provenance} to design proactive, on-chain compliance layers. Rather than relying on centralized entities to trace funds retroactively, or using fragile blocklists to catch non-compliant actors, \syss completely inverts the compliance model: it requires the user to proactively prove the legitimate origin (the provenance) of their funds prior to interaction with their target application. The core innovation of \syss enables users to construct and submit a zero-knowledge proof attesting to a continuous, compliant financial history---represented as a \emph{provenance sub-DAG} (Directed Acyclic Graph)---bridging their current funds back to a trusted, compliant origin set (\eg a KYC-compliant exchange). This yields three main architectural advantages:

\begin{itemize}
    \item[\ding{182}] \textbf{Proactive compliance with asymmetric computational scaling:} \syss outsources the computational burden of checking the historical graph traversal to the individual users (the prover). The smart contract acts merely as a verifier, confirming the zero-knowledge proof in $O(1)$ time and natively inheriting the base layer's security model.
    \item[\ding{183}] \textbf{Immunity to tainting attacks:} As the proof enforces that the deposited funds mathematically trace back to some compliant origin, the system is indifferent to how funds circulate or if an address has been passively ``tainted'' by malicious actors. If an honest user receives illicit funds, they simply exclude that specific transaction edge from their provenance DAG witness. 
    \item[\ding{184}] \textbf{Static over dynamic constraints:} By anchoring eligibility to rather static, compliant origins (such as regulated financial firms where users have already undergone KYC/AML), \syss removes the need for the verifier (\eg smart contract) to synchronize with rapidly changing blocklists/allowlists.
\end{itemize}

\subsection{Applications}

We present three promising on-chain applications of our \sys protocol across blockchain systems as follows. 

\pparagraph{Compliant Privacy Pools} 
Privacy pools (\eg Tornado Cash~\cite{pertsev2019tornado}) provide essential financial anonymity by severing the deterministic link between deposits and withdrawals. However, this obfuscation makes them highly attractive to malicious actors for laundering stolen cryptoassets, often leading to the entire pool being sanctioned by regulatory bodies (\eg OFAC~\cite{ofac-tornado-cash-2022}). Existing compliance solutions attempt to solve this via binary inclusion/exclusion lists~\cite{nadler2023tornado, baranski2023haze}, which suffer efficiency and usability issues. By deploying $\syss$ as a strict smart contract admission control mechanism, privacy pools can ensure that incoming deposits are exclusively funded by verified, compliant origins. The pool's anonymity set remains intact and perfectly untraceable, but the pool itself is guaranteed to be entirely free of illicit liquidity, resolving the fundamental tension between privacy and compliance.

\pparagraph{Proactive AML and Exchange Onboarding} Cryptocurrency exchanges (both centralized and decentralized) are mandated to conduct Anti-Money Laundering (AML) checks on client deposits~\cite{chainup_kyt_2026, azgad2023case}. Currently, this relies on continuous, retroactive chain-surveillance by third-party analytics firms---a highly expensive process that exposes exchanges to regulatory risk if illicit funds are identified \textit{after} they have entered the platform. To contextualize this financial burden, recent industry studies estimate that global financial institutions spend over \$206 billion annually on financial crime compliance, with AML operations consuming up to 19\% of a firm's annual revenue~\cite{lexisnexis_cost_2023, flagright_aml_2026}. For small-to-mid-sized cryptocurrency platforms alone, baseline KYC/AML overhead averages roughly \$620,000 per year, scaling rapidly into the tens of millions for major exchanges~\cite{ondato_kyc_2026}. $\syss$ completely inverts this model. By requiring a zero-knowledge provenance proof alongside the deposit transaction ($\tx_{\text{target}}$), exchanges can outsource the computational and analytical burden of AML to the user. The platform performs a proactive, $O(1)$ verification to guarantee the funds are clean before accepting them, drastically reducing compliance overhead while protecting the user's broader financial history from unnecessary corporate surveillance.

\pparagraph{Privacy-Preserving Undercollateralized Lending} 
The current Decentralized Finance (DeFi) lending ecosystem is highly capital-inefficient, structurally mandating severe overcollateralization (\eg $150\%$) because smart contracts inherently lack access to traditional credit-evaluation mechanisms and identities~\cite{binance_zkpass_2026}. Existing attempts to integrate off-chain credit scores via oracles introduce centralized trust assumptions and require users to completely forfeit their financial privacy to third-party data providers. $\syss$ provides the cryptographic foundation for native, trustless undercollateralized lending by treating proven historical liquidity as a proxy for creditworthiness. A prospective borrower can extract a sub-DAG demonstrating a cryptographically verified history of continuous, legitimate cash flows---such as regular deposits from an allowlisted set of employer addresses or consistent, liquidation-free interactions with other DeFi protocols---over a sustained period of time. By proving sufficient owned funds over time without leaking the user's total net worth, intermediate transaction topology, or specific counterparty identities, $\syss$ allows protocols to dynamically assess risk and offer undercollateralized loans while preserving user privacy.


\begin{figure}[t] 
    \centering
    \scalebox{0.85}{
        \begin{tikzpicture}[
            >=stealth,
            compliant/.style={circle, draw=green!60!black, fill=green!5, thick, minimum size=0.7cm, font=\scriptsize},
            dirty/.style={circle, draw=gray!60, fill=gray!10, dashed, thick, minimum size=0.7cm, font=\scriptsize, text=gray!70},
            addr/.style={circle, draw=blue!60!black, fill=blue!5, thick, minimum size=0.7cm, font=\scriptsize},
            ghostaddr/.style={circle, draw=gray!40, fill=gray!5, minimum size=0.7cm, font=\scriptsize, text=gray!50},
            target/.style={rectangle, rounded corners, draw=red!60!black, fill=red!5, thick, minimum height=0.6cm, minimum width=1.5cm, font=\scriptsize},
            noiseedge/.style={->, draw=gray!30, thin},
            pathedge/.style={->, draw=blue!70!black, ultra thick},
            labelstyle/.style={font=\tiny, text=black}
        ]

        \begin{scope}[shift={(0, 0)}]
            \node[font=\small\bfseries] at (1.5, 3.5) {Global Ledger State ($\mathcal{G}$)};
            
            \node[dirty] (D1) at (-2, 2.5) {$U_1$};
            \node[dirty] (D2) at (-2, 0.5) {$U_2$};
            \node[dirty] (D3) at (1, -2) {$U_3$};
            \node[dirty] (D4) at (3, 2.5) {$U_4$};
            
            \node[ghostaddr] (G1) at (0, 2) {};
            \node[ghostaddr] (G2) at (1, 1) {};
            \node[ghostaddr] (G3) at (3, -1.5) {};

            \node[compliant] (S1) at (-2, -1) {$S_1$};
            \node[compliant] (S2) at (-2, -2.5) {$S_2$};
            \node[addr] (A1) at (0, -1) {$A$};
            \node[addr] (A2) at (2, -0.5) {$B$};
            \node[target] (T1) at (4.5, -0.5) {$\tx_{\text{target}}$};

            \draw[noiseedge] (D1) -- (G1);
            \draw[noiseedge] (D2) -- (G1);
            \draw[noiseedge] (D2) -- (A1); 
            \draw[noiseedge] (G1) -- (G2);
            \draw[noiseedge] (G2) -- (A2); 
            \draw[noiseedge] (G2) -- (D4);
            \draw[noiseedge] (D3) -- (G3);
            \draw[noiseedge] (A1) -- (G3); 
            \draw[noiseedge] (G3) -- (T1);

            \draw[pathedge] (S1) -- (A1);
            \draw[pathedge] (S2) -- (A1);
            \draw[pathedge] (A1) -- (A2);
            \draw[pathedge] (A2) -- (T1);
        \end{scope}

        \draw[->, ultra thick, draw=black!70, dashed] (5.8, 0) -- node[above, font=\scriptsize\bfseries, align=center] {Retrospective\\Extraction} (8.5, 0);

        \begin{scope}[shift={(10, 0)}]
            \node[font=\small\bfseries] at (1.5, 3.5) {Proven Sub-DAG ($\mathcal{G}'$)};
            
            \node[compliant] (S1_ext) at (-1.5, 1) {$S_1$};
            \node[compliant] (S2_ext) at (-1.5, -2) {$S_2$};
            \node[addr] (A1_ext) at (0.5, -0.5) {$A$};
            \node[addr] (A2_ext) at (2.2, -0.5) {$B$};
            \node[target] (T1_ext) at (4.9, -0.5) {$\tx_{\text{target}}$};

            \draw[pathedge] (S1_ext) -- node[above, sloped, labelstyle] {$v_{e,1} \le v(\tx_1)$} node[below, sloped, labelstyle] {$h_1$} (A1_ext);
            \draw[pathedge] (S2_ext) -- node[above, sloped, labelstyle] {$v_{e,2} \le v(\tx_2)$} node[below, sloped, labelstyle] {$h_2$} (A1_ext);
            
            \draw[pathedge] (A1_ext) -- node[above, labelstyle] {$v_{e,3}$} node[below, labelstyle] {$h_3$} (A2_ext);
            
            \draw[pathedge] (A2_ext) -- node[above, labelstyle] {$v_{e,4} = V_{\text{total}}$} node[below, labelstyle] {$h_4$} (T1_ext);

            \node[font=\scriptsize, align=center, text=blue!80!black] at (0.5, -2.5) {\textbf{Flow Solvency:}\\$v_{e,3} \le v_{e,1} + v_{e,2}$};
            \node[font=\scriptsize, align=center, text=blue!80!black] at (2.5, 1) {\textbf{Strict Ordering:}\\$h_3 < h_4$};

        \end{scope}

        \matrix [
            at={(5, -4.5)}, 
            column sep=0.4cm, 
            row sep=0.2cm, 
            font=\scriptsize,
            ampersand replacement=\& 
        ] {
            \node[compliant, minimum size=0.4cm] {}; \& \node[anchor=west, inner sep=0] {Compliant Source ($\mathcal{C}$)}; \&
            \node[dirty, minimum size=0.4cm] {}; \& \node[anchor=west, inner sep=0] {Non-Compliant}; \&
            \draw[pathedge] (0,0) -- (0.6,0); \& \node[anchor=west, inner sep=0] {Proven Edge in $\mathcal{G}'$}; \\
            \node[addr, minimum size=0.4cm] {}; \& \node[anchor=west, inner sep=0] {Intermediate Address}; \&
            \node[target, minimum height=0.4cm, minimum width=0.8cm] {}; \& \node[anchor=west, inner sep=0] {Execution Payload}; \&
            \draw[noiseedge] (0,0) -- (0.6,0); \& \node[anchor=west, inner sep=0] {Ignored Edge}; \\
        };
        \end{tikzpicture}
    }
\caption{Compliant Subgraph Extraction. \textbf{(Left)} The global ledger $\mathcal{G}$ contains a co-mingled graph of compliant and unauthorized transactions. \textbf{(Right)} The prover utilizes a constrained graph search to extract a compliant sub-DAG $\mathcal{G}' \subseteq \mathcal{G}$. The nodal predicates (\eg Flow Solvency, Strict Temporal Ordering) are natively enforced over this isolated topology by the zero-knowledge circuit.}
    \label{fig:extraction_process}
\end{figure}
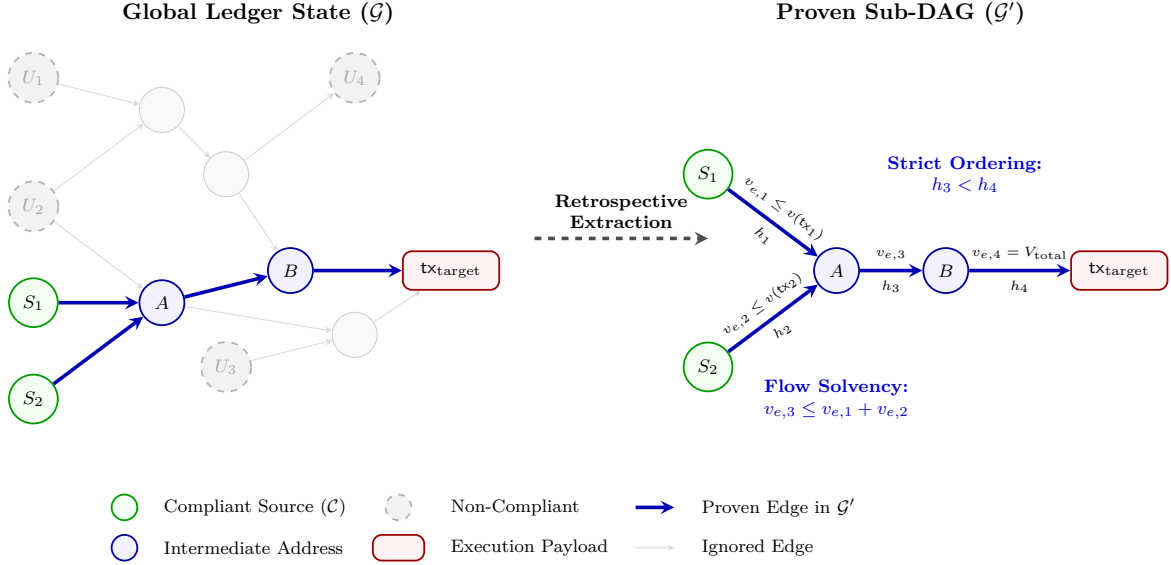

\subsection{Technical Overview}
\label{sec:technical_overview}

In this section, we provide a high-level technical description of our proposed provenance proof system, called \sys ($\syss$). Unlike prior approaches that rely on disjointed on-chain tracing or binary (non-)membership proofs in exclusion lists, our system introduces a unified cryptographic framework that enables users to extract a compliant on-chain sub-DAG and prove its integrity. 

\pparagraph{The Formal Provenance Model} 
An intuitive approach to generating a \syss is to first extract a sub-DAG $\mathcal{G}' \subseteq \mathcal{G}$ from the global transaction graph and enforce a set of predicates over its topology. However, the integrity of such a proof depends entirely on identifying a set of constraints that are both \textit{mutually exclusive} and \textit{collectively exhaustive} to protect against a diverse range of attack vectors. This necessitates a delicate identification of potential security violations as we address them in our core predicate design.

To prevent cyclic flow definitions and unify blockchain data structures, we define temporal nodes as $v = (a, h)$, where $a$ is an address and $h$ is a deterministic on-chain sequence coordinate. The (zero-knowledge) proof circuit should enforce a rigorous set of predicates to neutralize identified security threats:
\begin{description}
    \item[On-chain Integrity and Linkage] Every edge $e\in\mathcal{E}'$ must possess a valid inclusion proof demonstrating it is a finalized on-chain transaction. Furthermore, contiguous edges must align their sender and receiver addresses to prevent \emph{topology forgery}. We introduce a fractional capacity variable $v_e \le v(\tx)$, allowing a prover to attest to a specific portion of a transaction's value as compliant. This ensures that while a prover can extract fractional provenance, they cannot commit \emph{capacity forgery} by claiming compliant volumes that exceed the actual ledger records.
    \item[Source Anchoring] Every source node must possess a positive inclusion proof demonstrating membership in the compliant set $\mathcal{C}$, establishing a trust boundary for the entry of capital.
    \item[Temporal Ordering] This predicate prevents \emph{time-travel exploits}, where an adversary attempts to retroactively justify a historical withdrawal by citing a compliant inflow that occurred at a later date. By enforcing $h_v > h_u$ using sequence coordinates of the form $(\mathsf{height}, \mathsf{index})$ across every directed edge $(u, v)$, the circuit ensures that every value transfer follows the immutable arrow of time, precluding the use of future funds to legitimize past actions.
    \item[Flow Solvency and Distinctness] Every intermediate node satisfies Flow Solvency, meaning the total fractional capacity of all outgoing edges ($v_{\text{out}}$) must not exceed that of all incoming edges ($v_{\text{in}}$), \ie $\sum v_{\text{out}} \le \sum v_{\text{in}}$. Furthermore, we enforce \emph{(transaction) Distinctness} ($\tx_i \neq \tx_j$) across the sub-DAG to preclude \emph{intra-DAG path overlap}. This prevents an adversary from "recycling" a single compliant transaction to balance multiple outgoing branches within the same proof.
    \item[Terminal Execution] The proof is fused to the target state-transition ($\tx_{\text{target}}$). This prevents \emph{state-substitution} and \emph{replay attacks}, ensuring the proof is non-malleable and valid only for the intended depositor.
\end{description}

\pparagraph{Zero-Knowledge Proof Architecture}
Proving the satisfaction of these predicates over a transaction DAG natively would require a considerably large monolithic arithmetic circuit. To achieve concrete computational efficiency and $O(1)$ verification costs, our system leverages Incrementally Verifiable Computation (IVC) via modern folding schemes (\eg Nova~\cite{kothapalli2022nova}). 

Instead of generating a single proof for the entire graph, the prover computes a constant-size {step circuit} for each edge (verifying transaction inclusion and signatures) and a {merge circuit} for nodal intersections. These proofs are incrementally folded together as the prover traverses the DAG. Crucially, while the prover reconstructs the exact transaction path locally, the final $\syss$ submitted to the verifier is a zero-knowledge succinct argument. The proof asserts that the predicates hold for the target execution $V_{\total}$, but perfectly hides the intermediate flow of funds, the identities of intermediary addresses, and the specific origins utilized. This maintains the {verifiable cryptographic provenance} of the public ledger without enabling mass financial surveillance.

\pparagraph{Contributions} 
In summary, we make the following core contributions in this work: 

\begin{itemize}[leftmargin=*]
    \item \textbf{Formalization of Prover-Side Provenance:} We introduce \sys ($\syss$), formalizing a paradigm shift from centralized network/platform-side compliance surveillance to decentralized, user-side provenance. We establish a unified {temporal DAG} abstraction that captures both account-based and UTXO ledgers as a generalized value-flow model, underpinned by rigorous security definitions.

    \item \textbf{Efficient Protocol Instantiation:} We construct an efficient $\syss$ protocol utilizing Incrementally Verifiable Computation (IVC) to achieve constant-size, $O(1)$ on-chain verification. Users locally extract a compliant sub-DAG and prove (cryptographic) predicates, without leaking intermediate transaction topologies or specific origin addresses. Moreover, we investigate and propose robust application-layer defenses, including public nullifiers and private change commitments, to protect against double-spending of clean history.

    \item \textbf{Implementation and Empirical Evaluation:} We provide an Ethereum-compatible proof-of-concept of the $\syss$ framework. Our benchmarks demonstrate that continuous, proactive compliance is highly practical on consumer hardware, requiring only $\approx 1.8$\,s to incrementally update a user's proof via IVC. Furthermore, final on-chain verification is constant-time $\approx 1.5$\,ms ($\approx 800$k gas), regardless of the size of the compliant sub-DAG.
\end{itemize}

\section{Background}\label{sec:background}

\subsection{Basic Notations}\label{sec:notations}

The security parameter is denoted by $\lambda$.
Let $\mathsf{GroupGen}(\cdot)$ be a probabilistic polynomial-time (\textsf{PPT}) algorithm
that on input $1^\lambda$ outputs $(p, \mathbb{G}, g)$,
where $\mathbb{G}$ is a cyclic group of prime order $p$ generated by $g$.
We assume that the Decisional Diffie–Hellman (DDH) assumption holds in $\mathbb{G}$. We denote by $\mathbb{Z}_p$ the integers modulo $p$.
Sampling uniformly at random from a set $S$ is written as
$x \leftarrow_R S$.
We denote the negligible functions by $\mathsf{negl}(\lambda)$.

\subsection{Blockchain Transaction}\label{sec:blockchain_transaction}

\pparagraph{Account-Based and UTXO-based Models} Blockchains differ in how they represent state. In \emph{account-based} systems (\eg Ethereum~\cite{wood2014ethereum}),
each address maintains a balance, a nonce,
and transactions update account balances directly. In \emph{UTXO-based} systems (\eg Bitcoin~\cite{nakamoto2008bitcoin}),
transactions consume previously created unspent transaction outputs (UTXOs)
and create new transaction outputs,
with explicit value conservation at the transaction level, \ie sum of transaction inputs must be larger than or equal to the sum of the outputs. Thus, an account-based transaction graph may contain circles, while a UTXO-based transaction graph yields a directed acyclic graph. Despite these structural differences,
both models support the tracing of asset movements
through publicly recorded transaction histories.
Graph-based abstractions are commonly used
to reason about asset flow in both paradigms~\cite{kelen2025towards}.

\pparagraph{Transaction Graphs} Blockchains maintain an append-only public ledger
consisting of ordered transactions.
Each confirmed transaction records at least:
(i) a sender,
(ii) a receiver,
(iii) a transferred amount,
and (iv) a parent block  reference. Because transaction data is publicly available,
blockchains naturally admit a graph interpretation.
At a high level, one may view the ledger
as inducing a directed transaction graph $\mathcal{G} = (\mathcal{V}, \mathcal{E})$,
where vertices correspond to addresses
and edges correspond to asset transfers. This graph representation is widely used
in blockchain analytics and compliance systems
to study asset propagation and behavioral patterns~\cite{meiklejohn2013fistful}.
Temporal information provided by block inclusion
induces a time-respecting structure on this graph. This abstraction captures logical asset flow
independently of the underlying ledger implementation.
In UTXO-based systems, each consumed and created output
can be mapped to directed edges in this graph.
In account-based systems, balance updates are interpreted similarly.

\pparagraph{Ethereum} Ethereum has two types of accounts: (i) \emph{externally-owned} accounts (EOAs), controlled by a private key $\sk$ and identified by a public key $\pk$, and (ii) \emph{contract} accounts, controlled by their Ethereum Virtual Machine (EVM) (byte)code. The Ethereum blockchain tracks the state of every account~\cite{wood2014ethereum}. State changes are initiated through transactions coming from EOAs, which we simply refer to as accounts. A transaction consists of the destination account address, a signature $\sigma$, the transferred amount, an optional data field (representing inputs to a contract), a $\mathsf{gasLimit}$ value, and a $\mathsf{gasPrice}$ value. Every account is associated with a $\mathsf{nonce}$, a counter that is incremented by every \emph{transaction sent}. The signature $\sigma$ signs the aforementioned data with the sender's private key. During transaction processing, $\sigma$ is verified against the nonce value. Ethereum nodes keep track of the nonce for each account in their local state database, ensuring they know the next expected nonce for any transaction that the user sends. As a result, transactions cannot be “replayed” on the Ethereum network. An Ethereum address $\addr$ is deterministically derived from its corresponding public key $\pk$ after applying a cryptographic hash function.   

\pparagraph{Asset Tracing and Provenance} Given the public transaction graph $\mathcal{G} = (\mathcal{V}, \mathcal{E})$, asset tracing reduces to reachability analysis under temporal constraints. In account-based systems, balances are aggregated at the address level, and transactions update global account states. In UTXO-based systems, tracing follows consumption and creation of outputs. Under our abstraction, both models are represented uniformly as directed edges in $\mathcal{G}$. Tracing then consists of identifying time-respecting paths between addresses, possibly under additional constraints (\eg exclusion of labeled nodes). This graph-based view enables formal reasoning about asset provenance and compliance policies.

\subsection{Zero-Knowledge Argument of Knowledge}

A proof system enables a prover $\prover$ to convince a verifier $\verifier$ about some statement $u$ such that $\exists w: (u; w) \in \rel$, where $w$ is the corresponding witness and $\rel$ is a polynomial-time decidable relation. A proof of knowledge system further convinces the verifier that not only the witness exists, but also the prover knows it. When a proof system only demonstrates that some statement holds and does not leak any information about the witness, it is zero-knowledge. A proof system is an argument when (knowledge) soundness only holds against a computationally bounded prover under certain computationally hard assumptions. A proof is succinct if both the verification time and
the proof size grow polylogarithmically in the size of the computation. We present a formal definition for zero-knowledge succinct (non-interactive) argument of knowledge in \Cref{sec:zksnarks-formal}.


\section{System Model and Transaction Graph}
\label{sec:system-model}

We provide a formal abstraction of the underlying ledger to ensure that our proof system remains agnostic to the specific realization (\eg UTXO vs. account-based) of the underlying blockchain. We model the blockchain as a directed temporal asset transfer graph and explicitly state the assumptions 
under which our system is defined.

\begin{figure}[t]
\centering
\scalebox{0.5}{ 
\begin{tikzpicture}[
    >=Stealth,
    state/.style={rectangle, draw=black, thick, minimum width=2.8cm, minimum height=1.4cm, align=center, fill=blue!5},
    source/.style={rectangle, rounded corners, draw=green!60!black, fill=green!5, thick, align=center, minimum height=0.8cm},
    dirty/.style={rectangle, rounded corners, draw=gray, fill=gray!10, dashed, thick, align=center, minimum height=0.8cm},
    target/.style={rectangle, rounded corners, draw=red!60!black, fill=red!5, thick, align=center, minimum height=0.8cm},
    flow/.style={->, thick, draw=black!80},
    dflow/.style={->, dashed, thick, draw=gray},
    stateflow/.style={->, double, thick, draw=blue!60!black}
]

\node[state] (A1) at (0, 0) {Node $(a, h_1)$ \\ \small Compliant $V_1 = 10$};
\node[state] (A2) at (6, 0) {Node $(a, h_2)$ \\ \small Compliant $V_2 = 30$};
\node[state] (A3) at (12, 0) {Node $(a, h_3)$ \\ \small Compliant $V_3 = 30$};

\node[target] (Dest) at (17.5, 0) {Execution $\tx_{\text{target}}$};

\draw[stateflow] (A1) -- node[above] {\small Aggregation} (A2);
\draw[stateflow] (A2) -- node[above] {\small Maintenance} (A3);
\draw[flow] (A3) -- node[above, align=center] {\small Binding \\ $v = 25$} (Dest);

\node[source] (S1) at (-1.5, 3) {Source $S_1 \in \mathcal{C}$};
\node[source] (S2) at (4.5, 3) {Source $S_2 \in \mathcal{C}$};
\node[dirty] (U)  at (10.5, 3) {Source $U \notin \mathcal{C}$};

\draw[flow] (S1) -- node[left] {$+10$} (A1.north west);
\draw[flow] (S2) -- node[left] {$+20$} (A2.north west);
\draw[dflow] (U) -- node[right, align=left] {$+50$ \\ \small (Ignored)} (A3.north);

\draw[->, very thick, dashed, draw=gray] (-2.5, -2.0) -- (20, -2.0) node[right, text=black] {\textbf{Time (Sequence $h$)}};
\draw[gray, thick] (0, -1.8) -- (0, -2.2) node[below] {$h_1$};
\draw[gray, thick] (6, -1.8) -- (6, -2.2) node[below] {$h_2$};
\draw[gray, thick] (12, -1.8) -- (12, -2.2) node[below] {$h_3$};
\draw[gray, thick] (17.5, -1.8) -- (17.5, -2.2) node[below] {$h_4$}; 

\end{tikzpicture}
}
\caption{Account-Based Provenance Illustration. A single identity $a$ is mapped to a sequence of temporal nodes. The nodes act as an accumulated proven capacity, aggregating compliant value ($V_i$) from authorized sources ($\mathcal{C}$) across time ($h_1, h_2$). Crucially, unauthorized inflows at $h_3$ are excluded from the sub-DAG, maintaining isolation. The state is then atomically consumed by the {terminal execution binding} payload at $h_4$, satisfying fractional solvency ($25 \le 30$).} 
\label{fig:account_dag}
\end{figure}
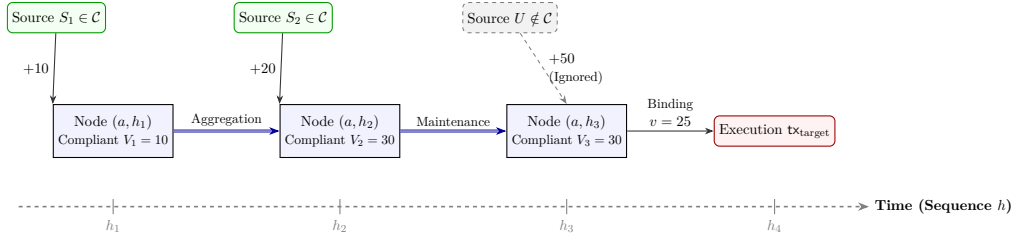

\subsection{The Global Ledger Abstraction}

Let $\mathcal{L}$ denote a distributed, append-only ledger maintained by a set of consensus participants. We model the ledger as an ordered sequence of blocks $\mathbf{B} = (B_0, B_1, \dots, B_n)$, where each block contains a header $H$ and an ordered sequence of atomic state transitions, or transactions.

\pparagraph{Identities and Value}
Let $\mathrm{Addr}$ be the set of cryptographic identities (\eg public keys or address hashes), and $\mathbb{N}$ denote the set of non-negative integers representing the value of some asset. The state of the ledger at any point defines a mapping that assigns value to identities, abstracting away the underlying accounting mechanism (\eg UTXO or account-based).

\pparagraph{Transaction Abstraction}
An abstract transaction $\tx$ is defined as a tuple:
$$
\tx = (s, r, v, h(\tx))
$$
where $s \in \mathrm{Addr}$ is the sender identity ($\mathsf{sender}(\tx)$), $r \in \mathrm{Addr}$ is the recipient identity ($\mathsf{receiver}(\tx)$), and $v \in \mathbb{N}$ is the transferred value ($v(\tx)$). 

Crucially, to guarantee a global ordering of all ledger events, let $h(\tx)$ denote the \textit{strict temporal coordinate} of the transaction, defined as the tuple $(\mathsf{height}, \mathsf{index})$, where $\mathsf{height}$ is the block number and $\mathsf{index}$ is the execution position of the transfer within that block. We assume the existence of a deterministic predicate $\mathsf{VerifyTxInclusion}(\tx, \Pi, B)$ which outputs $1$ if the atomic transfer $\tx$ is valid and correctly included in the block $B$ committed to by the canonical ledger.

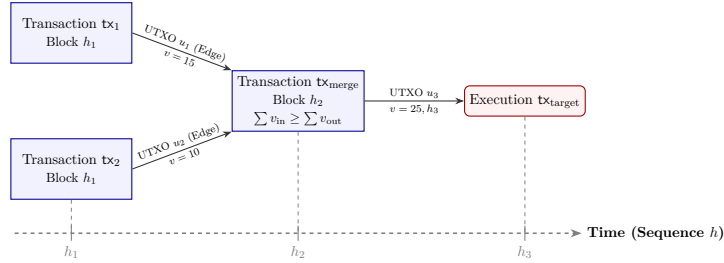
\begin{figure}[t]
\centering
\scalebox{0.5}{
\begin{tikzpicture}[
    >=Stealth,
    txnode/.style={rectangle, draw=blue!60!black, fill=blue!5, thick, minimum width=3.2cm, minimum height=1.6cm, align=center},
    target/.style={rectangle, rounded corners, draw=red!60!black, fill=red!5, thick, align=center, minimum height=0.8cm, minimum width=2.2cm},
    flow/.style={->, thick, draw=black!80},
    elab/.style={font=\scriptsize, inner sep=2pt}
]

\node[txnode] (tx1) at (0, 1.8) {Transaction $\mathsf{tx}_1$ \\ \small Block $h_1$};
\node[txnode] (tx2) at (0, -1.8) {Transaction $\mathsf{tx}_2$ \\ \small Block $h_1$};

\node[txnode] (txmerge) at (6, 0) {Transaction $\mathsf{tx}_{\text{merge}}$ \\ \small Block $h_2$ \\ \small $\sum v_{\text{in}} \ge \sum v_{\text{out}}$};

\node[target] (Dest) at (12, 0) {Execution $\mathsf{tx}_{\text{target}}$};

\draw[flow] (tx1.east) -- node[above, sloped, elab] {UTXO $u_1$ (Edge)} node[below, sloped, elab] {$v=15$} (txmerge.north west);
\draw[flow] (tx2.east) -- node[above, sloped, elab] {UTXO $u_2$ (Edge)} node[below, sloped, elab] {$v=10$} (txmerge.south west);

\draw[flow] (txmerge.east) -- node[above, elab] {UTXO $u_3$} node[below, elab] {$v = 25, h_3$} (Dest.west);

\draw[->, very thick, dashed, draw=gray] (-1.5, -3.5) -- (13.5, -3.5) node[right, text=black] {\textbf{Time (Sequence $h$)}};
\draw[gray, thick] (0, -3.3) -- (0, -3.7) node[below] {$h_1$};
\draw[gray, thick] (6, -3.3) -- (6, -3.7) node[below] {$h_2$};
\draw[gray, thick] (12, -3.3) -- (12, -3.7) node[below] {$h_3$};

\draw[dashed, gray] (tx2.south) -- (0, -3.5);
\draw[dashed, gray] (txmerge.south) -- (6, -3.5);
\draw[dashed, gray] (Dest.south) -- (12, -3.5);

\end{tikzpicture}
}
\caption{UTXO-Based Provenance Illustration. In the Bitcoin sub-DAG, \emph{transactions} act as the temporal nodes, while the discrete \emph{unspent outputs} (UTXOs) act as the directed edges. Flow Solvency natively resolves the multi-input aggregation ($15 + 10 \ge 25$) directly within the node $\mathsf{tx}_{\text{merge}}$ at block height $h_2$, securely linking the origin funds to the terminal execution payload at $h_3$.}
\label{fig:utxo_dag}
\end{figure}

\subsection{The Provenance DAG}

To reason about the origin of specific funds, we represent the ledger's history as a \textit{value-flow directed acyclic graph (DAG)}, denoted $\mathcal{G} = (\mathcal{V}, \mathcal{E})$. This mathematical structure naturally precludes circularity in provenance assertions—ensuring that account $A$ cannot be deemed compliant based on inflows from account $B$ if those funds originated from $A$ at a subsequent sequence coordinate. This ensures that every unit of compliant value can be traced back to an anchor in a globally defined \emph{compliant set} $\mathcal{C}$ (\eg a public registry or allowlist of authorized entities) through a finite, non-repeating history.\footnote{To focus on the core mechanics, we restrict all intermediate temporal nodes to Externally Owned Accounts (EOAs) and assume a single asset type (\eg native ETH).}

\pparagraph{Nodes (Temporal Identities)}
A node $n = (a, h) \in \mathrm{Addr} \times \mathcal{H}$ represents an address $a$ at a specific sequence coordinate $h$ within the sequence domain $\mathcal{H}$. Unlike standard transaction graphs where nodes are static addresses, this temporal tagging creates a unique ``snapshot'' of an identity at the exact moment of a state transition. 

\pparagraph{Edges (Compliant Flows)}
A directed edge $e = (n_u, n_v)$ exists between nodes $n_u = (a_u, h_u)$ and $n_v = (a_v, h_v)$ if there exists an on-chain transaction $\tx$ at sequence $h \in (h_u, h_v]$ that transfers value from $a_u$ to $a_v$. The edge is assigned a weight $v_e \le v(\tx)$, representing the fractional amount of compliant value passing between the nodes, as enforced by the underlying consensus rules and the circuit's predicates.

The fundamental security property of our model is \textit{flow solvency}. For any intermediate node $n \in \mathcal{V}$ (neither a source nor a terminal sink), let $\mathcal{E}_{\text{in}}(n)$ be the set of incoming edges and $\mathcal{E}_{\text{out}}(n)$ be the set of outgoing edges. The graph is valid if and only if every intermediate node satisfies:
$$
\sum_{e_{\text{out}} \in \mathcal{E}_{\text{out}}(n)} v_{e_{\text{out}}} \le \sum_{e_{\text{in}} \in \mathcal{E}_{\text{in}}(n)} v_{e_{\text{in}}}
$$
This constraint treats compliant capacity as a finite, non-inflationary resource. 

\pparagraph{Anchors and Sinks}
A provenance proof $\pi_{\ori}$ extracts a compliant sub-DAG $\mathcal{G}' \subseteq \mathcal{G}$ satisfying the following main conditions:
\begin{enumerate}[leftmargin=*]
    \item \textbf{Source Anchoring:} Every source node  in $\mathcal{G}'$ must correspond to an address $a \in \mathcal{C}$ (the compliant set defined above) at some sequence $h$. 
    \item \textbf{Terminal Execution Binding:} The sink/target node $n_{\text{sink}}$ in $\mathcal{G}'$ is defined by the target execution payload $\tx_{\text{target}}$. The graph must terminate at the exact sequence coordinate $h_{\text{target}}$ associated with the on-chain state-transition.
    \item \textbf{Value Sufficiency:} The sum of compliant values reaching the sink node must satisfy the payload's required volume $V_{\total}$:
    $$
    \sum_{e \in \mathcal{E}'_{\text{in}}(n_{\text{sink}})} v_e \ge V_{\total}
    $$
\end{enumerate}

\pparagraph{Ledger Versatility}
By parameterizing nodes as temporal pairs $(a, h)$, the model unifies heterogeneous ledgers into a single resource-solvency framework. In a UTXO paradigm, nodes represent discrete transactions executed at coordinate $h$, and the directed edges denote the unspent outputs (UTXOs) transferring value between them. In an account-based paradigm, nodes capture an identity's accumulated proven capacity at a specific execution time. In both cases, the solvency requirement ensures that outgoing compliant transfers are strictly bounded by verified receipts. For the sake of notational clarity and without loss of generality, we focus on an account-based paradigm for the remainder of the paper.

\section{\sys} 

\subsection{Proof System for On-Chain Provenance Flow}

In this section, we define the \sys (\syss) system and present its formal definitions and security properties. Our approach models on-chain provenance as a flow of ``compliant value'' through a transaction graph, enabling a prover to demonstrate that a target execution payload $\tx_{\text{target}}$ is funded by a total compliant value $V_{\total}$ originating from a set of publicly committed source addresses $\mathcal{C}$. The system preserves the privacy of intermediate transaction data and the specific identities of the source addresses while providing a succinct verification of the total value's legitimacy.

\pparagraph{Parameters and Commitments} 
We assume the underlying distributed ledger maintains a canonical chain of blocks, denoted as $H$. Each finalized block $B \in H$ cryptographically commits to a transaction set, supporting succinct inclusion proofs $\Pi$ that attest a specific transaction $\tx$ was successfully included and executed in $B$. Let $c_{\mathcal{C}}$ denote a public cryptographic commitment (\eg a Merkle root) to the set $\mathcal{C} \subseteq \mathrm{Addr}$ of compliant addresses. Let $R_H$ denote a public commitment to the canonical chain head (\eg the state root of the latest block in $H$). Let $\tx_{\text{target}}$ be the specific execution payload of the target application being funded, and $V_{\total} \in \mathbb{N}$ be the total amount of compliant value being justified.

\pparagraph{Statements and Witnesses}
A public \syss statement is defined as:
$$
x = (c_{\mathcal C},\, R_H,\, \tx_{\text{target}},\, V_{\total}).
$$
A witness $w$ is a provenance DAG $\mathcal{G}' = (\mathcal{V}', \mathcal{E}')$ alongside a set of corresponding state inclusion proofs. In this graph topology, aligning with our account-based provenance model (see~\Cref{fig:account_dag}), the vertices $\mathcal{V}'$ represent \emph{temporal nodes} $n = (a, h)$, and the directed edges $\mathcal{E}' \subset \mathrm{Tx}$ represent on-chain transactions transferring value in between. Each source node $n_{\text{source}} \in \mathcal{V}'$ must be a member of the compliant set $\mathcal{C}$. Each edge $e \in \mathcal{E}'$ must be accompanied by its cryptographic inclusion proof $\Pi$ and corresponding block $B$ on the canonical chain.

\subsection{Formal Predicates}\label{sec:formal_predicates}
To guarantee the cryptographic integrity of the provenance sub-DAG $\mathcal{G}' = (\mathcal{V}', \mathcal{E}')$, the zero-knowledge circuit must strictly enforce the following set of predicates for every submitted proof $\pi_{\ori}$. Let $h(\tx)$ denote the sequence index (\eg block height) of transaction edge $\tx$, and $v(\tx)$ denote its transferred value. Define the predicate $\mathsf{Prov}(x,w) \in \{0,1\}$ for a statement $x$ and witness $w$ to output $1$ if and only if all the following conditions hold:

\begin{enumerate}[label=(\roman*)]
  \item \textbf{On-Chain Integrity and Distinctness:} For every edge $e \in \mathcal{E}'$ in the DAG claiming a compliant value $v_e$, the prover provides a valid inclusion proof $\Pi$ mapping the edge to a finalized on-chain transaction $\tx$ and block $B$ such that:
  $$
    \mathsf{VerifyTxInclusion}(\tx,\Pi,B)=1 \quad \text{and} \quad \mathsf{VerifyChainToHead}(B, H)=1.
  $$
  Furthermore, the relation enforces the following across the entire sub-DAG:
  $$
    v_e \le v(\tx) \quad \text{and} \quad \forall e_i, e_j \in \mathcal{E}' \text{ where } i \neq j, \ \tx_i \neq \tx_j.
  $$
  This condition binds the abstract provenance graph to the on-chain ledger. It guarantees that every edge corresponds to a finalized canonical transaction, permits fractional capacity extraction from blended transactions ($v_e \le v(\tx)$), and precludes any exploits where a prover feeds the same transaction into multiple branches.\medskip

\item \textbf{Source Anchoring:} For every source node $n_{0,j} = (a_{0,j}, h_{0,j}) \in \mathcal{V'}$ (\ie temporal nodes with an in-degree of zero in the extracted sub-DAG), the associated address $a_{0,j}$ must be a member of the compliant set:
  $$ 
    \mathsf{VerifyMember}(a_{0,j},\pi_{\mathcal C},c_{\mathcal C})=1.
  $$
    This predicate defines the trust boundary of the system. It enforces that every unit of compliant value enters the graph through an address $a_{0,j}$ that is a member of the set $\mathcal{C}$. To achieve this, the on-chain commitment $c_{\mathcal{C}}$ must be instantiated as a \emph{positive accumulator} (\eg a Merkle tree root) that supports efficient, verifiable inclusion proofs. The membership proof $\pi_{\mathcal{C}}$ relative to this accumulator ensures that the prover cannot inject value from unauthorized or unknown sources into the flow.\medskip

    \item \textbf{Edge Linkage:} For every pair of adjacent edges $e_{\text{in}} = (n_u, n_v)$ and $e_{\text{out}} = (n_v, n_w)$ in $\mathcal{E}'$ connected by a shared intermediate temporal node $v \in \mathcal{V}'$, the circuit must verify that the cryptographically recovered destination identity of the incoming transfer perfectly matches the verified source identity of the outgoing transfer: 
  $$
    \mathsf{receiver}(\tx_{e_{\text{in}}}) = \mathsf{sender}(\tx_{e_{\text{out}}}).
  $$
  This explicit continuity constraint prevents \textit{topology forgery}. Without it, a malicious prover could stitch together completely disconnected, valid on-chain transactions to fabricate a fraudulent path between the compliant set $\mathcal{C}$ and the target execution payload, circumventing the causal flow of funds. \medskip

  \item \textbf{Temporal Ordering:} For every pair of adjacent edges $e_{\text{in}}$ and $e_{\text{out}}$ in $\mathcal{E}'$ intersecting at a shared intermediate node $v \in \mathcal{V}'$, the relation enforces that the incoming transaction strictly precedes the outgoing transaction:
  $$
    h(e_{\text{in}}) < h(e_{\text{out}}).
  $$
  This condition prevents chronological manipulation. Without strict acyclicity, an adversary could retroactively exploit future compliant inflows to artificially legitimize historically non-compliant transfers, circumventing temporal causality.\medskip


\item \textbf{Flow Solvency:} For every intermediate address $a$ in the DAG, let $V_{\text{in}}(a) = \sum_{e \in \mathcal{E}'_{\text{in}}(a)} v_e$ be the sum of compliant fractional edge values received, and $V_{\text{out}}(a) = \sum_{e \in \mathcal{E}'_{\text{out}}(a)} v_e$ be the sum of values transferred. The relation enforces:
  $$
    V_{\text{out}}(a) \le V_{\text{in}}(a).
  $$
  This condition ensures that compliant flow is never artificially amplified and that any outgoing transfer is fully backed by verified incoming funds.\medskip

\item \textbf{Terminal Execution Binding:} The final edge in the provenance graph must deterministically bind to the exact real transaction payload ($\tx_{\text{target}}$) being executed. The aggregate compliant value $V_{\text{final}}$ reaching this terminal edge satisfies:
  $$
    \mathsf{VerifyExecutionBinding}(\tx_{\text{target}}) = 1 \quad \text{and} \quad V_{\text{final}} \ge V_{\total}.
  $$
This final check ensures that the aggregate compliant value successfully funds the designated public target $V_{\total}$. While the application layer handles the state-management of the execution (\eg nullifiers), the mathematical abstraction of the DAG requires a definitive algorithmic sink.\footnote{Intuitively, a cryptographic protocol designed for authorization must bind the proof of authorization to the specific payload being authorized.} Critically, by requiring $\mathsf{VerifyExecutionBinding}$ natively within the core relation, the circuit atomically fuses the verified compliant sub-DAG directly to the on-chain state-transition.
\end{enumerate}

Putting all predicates together, we have a complete, unbroken chain: \ding{182} Integrity (it happened), \ding{183} Anchoring (it started compliant), \ding{184} Linkage (it connects), \ding{185} Acyclicity (it moves forward in time), \ding{186} Solvency (it does not inflate), and \ding{187} Binding (it authorizes the specific action).

\begin{figure}[t]
    \centering
    \scalebox{0.85}{
        \begin{tikzpicture}[
            >=stealth,
            every node/.style={align=center, font=\footnotesize},
            source/.style={circle, draw=green!60!black, fill=green!5, thick, minimum size=0.9cm, inner sep=1pt},
            addr/.style={circle, draw=blue!60!black, fill=blue!5, thick, minimum size=0.9cm, inner sep=1pt},
            target/.style={rectangle, rounded corners, draw=red!60!black, fill=red!5, thick, minimum height=0.6cm, minimum width=2cm},
            dirty/.style={circle, draw=gray, fill=gray!10, dashed, thick, minimum size=0.9cm, inner sep=1pt},
            edge/.style={->, thick, draw=black!80},
            ignored/.style={->, dashed, draw=gray, thick},
            elab/.style={font=\scriptsize, inner sep=2pt} 
        ]

        \begin{scope}[shift={(-3.8, 0)}]
            \node (title1) at (0,0) {\textbf{Scenario 1: Single Path}};
            
            \node[addr] (a1) [below=0.6cm of title1] {$A$};
            \node[source] (s1) [left=1.3cm of a1] {$S_1 \in \mathcal{C}$};
            \node[target] (t1) [right=1.3cm of a1] {$\tx_{\text{target}}$};

            \draw[edge] (s1) -- node[above, elab] {$v=100$} node[below, elab] {$h_1$} (a1);
            \draw[edge] (a1) -- node[above, elab] {$v=100$} node[below, elab] {$h_2$} (t1);
        \end{scope}

        \begin{scope}[shift={(3.8, 0)}]
            \node (title2) at (0,0) {\textbf{Scenario 2: Multi-Source Flow}}; 
            
            \node[addr] (a2) [below=0.8cm of title2] {$B$};
            \node[source] (s2) [left=1.3cm of a2, yshift=0.8cm] {$S_2 \in \mathcal{C}$};
            \node[source] (s3) [left=1.3cm of a2, yshift=-0.8cm] {$S_3 \in \mathcal{C}$};
            \node[target] (t2) [right=1.2cm of a2] {$\tx_{\text{target}}$};

            \draw[edge] (s2) -- node[above, sloped, elab] {$v=80$} node[below, sloped, elab] {$h_3$} (a2);
            \draw[edge] (s3) -- node[above, sloped, elab] {$v=50$} node[below, sloped, elab] {$h_4$} (a2);
            \draw[edge] (a2) -- node[above, elab] {$v=100$} node[below, elab] {$h_5$} (t2);
        \end{scope}

        \begin{scope}[shift={(0, -3.8)}] 
            \node (title3) at (0,0) {\textbf{Scenario 3: Complex DAG}};
            
            \node[dirty] (u1) [below=0.3cm of title3] {$U \notin \mathcal{C}$};
            \node[addr] (a3) [below=0.9cm of u1] {$C$};
            
            \node[addr] (a4) [left=1.3cm of a3, yshift=1.2cm] {$D$};
            \node[addr] (a5) [left=1.3cm of a3, yshift=-1.2cm] {$E$};
            
            \node[source] (s4) [left=1.3cm of a4] {$S_4 \in \mathcal{C}$};
            \node[source] (s5) [left=1.3cm of a5] {$S_5 \in \mathcal{C}$};
            \node[target] (t3) [right=1.3cm of a3] {$\tx_{\text{target}}$};
            
            \draw[edge] (s4) -- node[above, elab] {$v=50$} node[below, elab] {$h_6$} (a4);
            \draw[edge] (s5) -- node[above, elab] {$v=50$} node[below, elab] {$h_7$} (a5);
            \draw[edge] (a4) -- node[above, sloped, elab] {$v=50$} node[below, sloped, elab] {$h_8$} (a3);
            \draw[edge] (a5) -- node[above, sloped, elab] {$v=50$} node[below, sloped, elab] {$h_9$} (a3);
            \draw[edge] (a3) -- node[above, elab] {$v=100$} node[below, elab] {$h_{11}$} (t3);
            
            \draw[ignored] (u1) -- node[right, elab] {$v=500, h_{10}$} (a3);
        \end{scope}

        \matrix [
            at={(0, -9.5)}, 
            column sep=0.5cm, 
            row sep=0.3cm, 
            font=\scriptsize,
            ampersand replacement=\& 
        ] {
            \node[source, minimum size=0.4cm] {}; \& \node[anchor=west, inner sep=0] {Compliant Source}; \&
            \node[addr, minimum size=0.4cm] {}; \& \node[anchor=west, inner sep=0] {Intermediate Address}; \&
            \node[target, minimum height=0.4cm, minimum width=0.8cm] {}; \& \node[anchor=west, inner sep=0] {Execution Payload}; \\
            
            \node[dirty, minimum size=0.4cm] {}; \& \node[anchor=west, inner sep=0] {Non-Compliant Source}; \&
            \draw[edge] (0,0) -- (0.6,0); \& \node[anchor=west, inner sep=0] {Proven Edge in $\mathcal{G}'$}; \&
            \draw[ignored] (0,0) -- (0.6,0); \& \node[anchor=west, inner sep=0] {Ignored Edge (Excluded)}; \\
        };

        \end{tikzpicture}
    }
\caption{Prover's Sub-DAG Witness ($\mathcal{G}'$). Scenario 1 isolates a single path via greedy search. Complex topologies may require capacity-balanced sub-DAGs: Scenario 2 demonstrates fractional consumption ($v_{\text{out}} \le \sum v_{\text{in}}$) absorbing surplus capacity, while Scenario 3 excludes non-compliant inflows (dashed gray) for strict Source Anchoring. Temporal acyclicity ($h_i < h_{i+1}$) is strictly enforced across all valid edges.}
    \label{fig:dag_scenarios}
\end{figure}
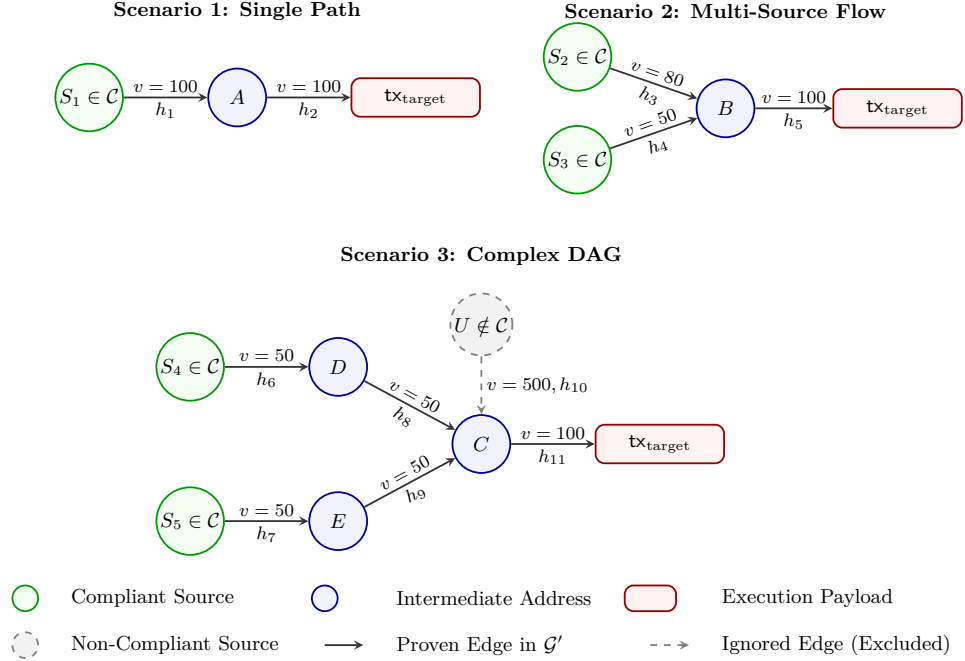

\begin{definition}[\sys (\syss)]
An on-chain proof system for (recursive) provenance consists of a tuple of \textsf{PPT} algorithms $(\syss.\mathsf{Setup},\syss.\mathsf{Prove},\syss.\mathsf{Verify})$ defined as follows.
\end{definition}

\begin{itemize}[leftmargin=*]
  \item \textbf{\syss.\textsf{Setup}}$(1^{\lambda}) \rightarrow (pp)$.
  \emph{On input security parameter $1^\lambda$, outputs public parameters $pp$, including a Common Reference String ($\crs$) for the underlying proof system, and an initial state parameter $(c_{\mathcal C},\, R_H,\, \mathsf{aux})$. Here, $c_{\mathcal C}$ commits to the compliant set $\mathcal{C} \subseteq \mathrm{Addr}$, $R_H$ denotes the canonical chain-head commitment, and $\mathsf{aux}$ contains auxiliary system parameters (\eg supported asset identifiers).}

  \item \textbf{\syss.\textsf{Prove}}$(pp, x, w) \rightarrow \pi_{\ori}$.
  \emph{On input public parameters $pp$, and public instance $x := (c_{\mathcal C},\, R_H,\, \tx_{\text{target}},\, V_{\total})$, and a provenance DAG witness $w$, outputs a proof $\pi_{\ori}$. This proof attests that there exists a valid witness $w$ such that $(x,w) \in \mathsf{Prov}$ for the public statement $x := (c_{\mathcal C},\, R_H,\, \tx_{\text{target}},\, V_{\total})$.}

  \item \textbf{\syss.\textsf{Verify}}$(pp, x, \pi_{\ori}) \rightarrow \{0,1\}$.
  \emph{Deterministically outputs $1$ iff $\pi_{\ori}$ is a valid, succinct proof for the public instance $x$. The verification cost is strictly sublinear (independent of the depth or breadth of the provenance DAG $w$), ensuring that the aggregate compliant value routed to $\tx_{\text{target}}$ is at least $V_{\total}$ and satisfies all required predicates.}
\end{itemize}

\subsection{Formal Security Definitions and Properties}

The non-interactive, prover-side nature of $\syss$ allows us to establish security via a rigorous property-based reduction. In our model, the canonical blockchain serves as the trusted ledger functionality. We define the protocol's security by ensuring that any extracted witness satisfying our provenance relation $R_{\syss}$ inherently guarantees three fundamental dimensions of ledger-based integrity: 
\begin{enumerate}
    \item \textbf{Existence:} Enforced via On-chain Integrity and Source Anchoring.
    \item \textbf{Continuity:} Enforced via Edge Linkage, Temporal Ordering, and Flow Solvency.
    \item \textbf{Non-Reusability:} Enforced via (transaction) Distinctness and Terminal Execution bounds.\footnote{We stress that this property is local to a single proof instance; preventing replay or double-spending of compliant history across multiple proofs requires stateful application-layer defenses, such as nullifiers and change commitments, which we discuss in~\Cref{sec:double_spend_history}.}
\end{enumerate}

We refer the reader to~\Cref{tab:threat_model} for a comprehensive mapping of these dimensions to their respective predicates and mitigated attack vectors. 

To formalize this, let $\syss = (\syss.\mathsf{Setup}, \syss.\mathsf{Prove}, \syss.\mathsf{Verify})$ be a non-interactive argument of knowledge for the provenance relation $R_{\syss}$, with the associated language $L_{\syss} := \{ x \mid \exists w : (x,w) \in R_{\syss} \}$. Public statements are of the form $x = (c_{\mathcal{C}}, R_H, \tx_{\text{target}}, V_{\total})$. The foregoing protocol dimensions reduce to the following formal cryptographic properties of the \syss proof system.

\begin{itemize}[leftmargin=*]
    \item \textbf{Completeness.} 
    The proof system is \emph{complete} if an honest prover with a valid witness can always convince an honest verifier. For every security parameter $\lambda$ and every $(x,w) \in R_{\syss}$:
    $$
    \Pr \left[ \syss.\mathsf{Verify}(pp, x, \pi) = 1 : 
    \begin{matrix} 
    (pp) \leftarrow \syss.\mathsf{Setup}(1^\lambda) \\ 
    \pi \leftarrow \syss.\mathsf{Prove}(pp, x, w) 
    \end{matrix} 
    \right] \ge 1 - \mathsf{negl}(\lambda).
    $$

    \item \textbf{(Computational) Soundness.} 
    The proof system is \emph{sound} if a malicious prover cannot forge a proof for a false statement. For every \textsf{PPT} adversary $\mathcal{A}$:
   $$
    \Pr \left[ \syss.\mathsf{Verify}(pp, x, \pi) = 1 : 
    \begin{matrix} 
    (pp) \leftarrow \syss.\mathsf{Setup}(1^\lambda) \\ 
    (x, \pi) \leftarrow \mathcal{A}(pp) \\ 
    x \notin L_{\syss} 
    \end{matrix} 
    \right] \le \mathsf{negl}(\lambda).
    $$

    \item \textbf{Knowledge Soundness (Extractability).} 
    The proof system satisfies \emph{knowledge soundness} if whenever an adversary produces a valid proof, it must ``know'' the corresponding witness. Formally, for every \textsf{PPT} adversary $\mathcal{A}$ taking auxiliary input $z$, there exists a \textsf{PPT} extractor $\mathcal{E}_{\mathcal{A}}$ such that:
    $$
    \Pr \left[ 
    \begin{matrix} 
    \syss.\mathsf{Verify}(pp, x, \pi) = 1 \\ 
    \land \ (x, w) \notin R_{\syss} 
    \end{matrix} : 
    \begin{matrix} 
    (pp, st_0) \leftarrow \syss.\mathsf{Setup}(1^\lambda) \\ 
    (x, \pi) \leftarrow \mathcal{A}(pp, z) \\ 
    w \leftarrow \mathcal{E}_{\mathcal{A}}(pp, z) 
    \end{matrix} 
    \right] \le \mathsf{negl}(\lambda).
    $$
    \emph{This standard extractability formulation ensures that any prover capable of producing a valid provenance certificate must necessarily possess the complete underlying DAG witness, including the valid transaction inclusion proofs, the valid Temporal Ordering, and the exact summation trace.} \medskip

    \item \textbf{Zero-Knowledge.} 
    The proof system is \emph{computational zero-knowledge} if the proof leaks no information about the private witness $w$. Formally, there exists a PPT simulator $(\syss.\mathsf{SimSetup}, \break \syss.\mathsf{SimProve})$ such that for all \textsf{PPT} distinguishers $\mathcal{D}$ and all $(x,w) \in R_{\syss}$ it holds:
{\small
$$
\begin{aligned}
\Bigg| &\Pr \Bigg[ \mathcal{D}(pp, \pi) = 1 : \begin{matrix} (pp) \leftarrow \syss.\mathsf{Setup}(1^\lambda) \\ \pi \leftarrow \syss.\mathsf{Prove}(pp, x, w) \end{matrix} \Bigg] \\
- &\Pr \Bigg[ \mathcal{D}(pp', \pi') = 1 : \begin{matrix} (pp', \tau) \leftarrow \syss.\mathsf{SimSetup}(1^\lambda) \\ \pi' \leftarrow \syss.\mathsf{SimProve}(pp', \tau, x) \end{matrix} \Bigg] \Bigg| \le \mathsf{negl}(\lambda).
\end{aligned}
$$
}
    \emph{Here, $\tau$ represents the simulation trapdoor. This property guarantees that the public instance $x$ and the proof $\pi$ reveal nothing about the intermediary transactions, the specific origins utilized, or the network topology of the sub-DAG.}
\end{itemize}

\section{Realizing \texorpdfstring{$R_{\syss}$}{R_prov} over Ethereum}
\label{sec:realization-ethereum}

We present a realization of the recursive provenance relation $R_{\syss}$ tailored for the Ethereum state model. Implementing these predicates requires a choice between two primary cryptographic architectures:

\begin{itemize}
    \item \textbf{Monolithic SNARKs (\eg Groth16):} In this approach, the entire provenance sub-DAG is unrolled into a single, static arithmetic circuit. While conceptually simple, the prover's memory requirements scale linearly with the number of transactions, making it prohibitive for deep histories or complex branching.
    \item \textbf{Folding-based IVC (\eg Nova):} By leveraging Incrementally Verifiable Computation (IVC), we can ``fold'' each transaction edge or nodal merge into a constant-size accumulator. This allows for scalable proof generation for arbitrary DAG depths while maintaining a constant-size on-chain verification cost.
\end{itemize}

Given the dynamic nature of on-chain transaction history, we instantiate our system using the Nova folding scheme.

\subsection{Illustrative Example: Aggregation and Binding}
To illustrate the mechanics of our predicates, consider a target execution $\tx_{\text{target}}$ requiring $V_{\total} = 25$. Let the committed source set $\mathcal{C}$ include entities $S_1$ and $S_2$.
\begin{enumerate}[leftmargin=*]
    \item \textbf{Compliant Aggregation:} Address $A$ receives two distinct transfers from the compliant set: $\tx_1 = (S_1, A, 10, {h}_1)$ and $\tx_2 = (S_2, A, 20, {h}_2)$. The prover generates a certificate attesting to a total incoming flow of $30$ units.
    \item \textbf{Exclusion of Non-Compliant Funds:} Address $A$ also receives $50$ units from an unauthorized source $U \notin \mathcal{C}$ via $\tx_3$. While the on-chain balance is $80$, the prover omits $\tx_3$ from the witness. The extracted sub-DAG isolates only the $30$ verified units.
    \item \textbf{Fractional Execution:} At coordinate ${h}_4$ (where ${h}_4 > \max({h}_1, {h}_2)$), address $A$ initiates $\tx_{\text{target}}$ for $25$ units. The circuit confirms $25 \le 30$ and binds the proof to the payload.
\end{enumerate}

The resulting proof $\pi_{\ori}$ demonstrates that the $25$ units are supported by the continuous provenance of $S_1$ and $S_2$, resolving fractional consumption within the graph topology.

\subsection{Realization of the Nodal Predicates} \label{sec:nodal_realization}
The core realization is an augmented arithmetic circuit $F'$ that executes the step-wise state transition. The IVC state at step $i$ is defined as $st_i = (a_i, V_i, {h}_i, \mathcal{H}_i)$, where $a_i$ is the destination address, $V_i$ is the accumulated value, ${h}_i = (\mathsf{height}, \mathsf{index})$ is the strict temporal coordinate, and $\mathcal{H}_i$ is a rolling commitment to the set of transaction identifiers included in the sub-path.

\pparagraph{Canonicality and Execution Check} 
To avoid the prohibitive cost of verifying the entire consensus mechanism (\eg Ethash or Casper) in-circuit, the system utilizes recursive inclusion verification. Each step $i$ provides an inclusion witness (\eg a Merkle--Patricia Trie proof) against a block header's \texttt{transactionsRoot}, an execution witness against its \texttt{receiptsRoot} to guarantee the transaction successfully resolved (status $\neq 0$), and a parent-chain witness leading to the globally finalized checkpoint $B_{\text{recent}}$. The circuit enforces:
$$
\mathsf{VerifyTx}(\tx_i, B_i.\mathsf{txRoot}) \land \mathsf{VerifyReceipt}(\tx_i, B_i.\mathsf{receiptRoot}) \land \dots \land B_{\text{chain}} = B_{\text{recent}}
$$
This dynamically induces a cryptographic chain of trust---proving inclusion, successful EVM execution, and canonical finality---from the specific transaction edge directly to the finalized global state.\footnote{The EVM only provides the most recent 256 block headers. However, verifying deep historical transactions does not introduce new trust assumptions. For older transactions, the prover fetches historical checkpoints off-chain from standard Ethereum data availability layers (\eg archive nodes). The circuit cryptographically bridges the deep historical target block to its closest checkpoint, which is then trustlessly authenticated against an on-chain historical state cache (\eg a Merkle Mountain Range~\cite{bonneau2025merkle} maintained by infrastructure like Axiom~\cite{axiom_v1}). As Ethereum relies on these historical checkpoints to sync nodes and prevent long-range attacks, our system inherits the base layer's existing security model.}

\pparagraph{Field-Bounded Conservation and Distinctness}
The circuit evaluates flow strictly through bounded summation gates to prevent prime-field overflow. Because zero-knowledge inequality checks require safe bit-widths, values are constrained to $2^{252}$ (safely within the SNARK scalar field $\mathbb{F}_p$):
\begin{itemize}
    \item \textbf{Transfer Step:} For an outgoing transaction $v_{\text{out}}$, the circuit verifies the capacity $v_{\text{out}} \le V_i$ alongside the bit-width bounds $v_{\text{out}} < 2^{252}$ and $V_i < 2^{252}$ to guarantee safe arithmetic comparisons.
    \item \textbf{Merge Step:} When aggregating proofs $st_{i,1}$ and $st_{i,2}$ at a node, the circuit enforces \emph{(transaction) Distinctness} to prevent double-counting edges:
    $$
    V_{\text{merged}} = V_{i,1} + V_{i,2} \quad \text{s.t.} \quad V_{\text{merged}} < 2^{252} \wedge \mathcal{H}_{i,1} \cap \mathcal{H}_{i,2} = \emptyset
    $$
\end{itemize}

\pparagraph{Strict Temporal Acyclicity}
The circuit enforces chronological succession using lexicographical comparison on the ${h} = (\mathsf{height}, \mathsf{index})$ tuple:
\begin{itemize}
    \item \textbf{Transfer Step:} ${h}_i < {h}(\tx_{\text{out}})$.
    \item \textbf{Merge Step:} ${h}_{\text{merged}} = \max({h}_{i,1}, {h}_{i,2})$.
\end{itemize}

\pparagraph{Non-Native ECDSA Recovery}
To satisfy \emph{Edge Linkage}, we perform $s = \mathsf{ecrecover}(\tx)$ in-circuit. Using non-native arithmetic gadgets, we ensure the recovered address is cryptographically bound to the transaction, preventing provers from fabricating source identities.

\pparagraph{MPT and RLP Extraction Constraints}
The circuit RLP-decodes the block header to extract the block number and \texttt{transactionsRoot}. For transaction decoding, we utilize bit-width boundary gadgets ($< 2^{252}$) immediately upon extracting the value field to ensure arithmetic safety before any downstream evaluation.

\subsection{Discussion}

\pparagraph{Preventing Topology Forgery}
A fundamental security requirement of the provenance sub-DAG $\mathcal{G}'$ is the preservation of a strict, unbroken cryptographic chain of custody. If the recursive circuit merely verifies the $\mathsf{receiver}(\tx_i)$ of an incoming edge while assuming the authorization of the subsequent outgoing edge $\tx_{i+1}$, the system becomes vulnerable to \textit{topology forgery}. In this exploit, a malicious prover stitches together disparate transaction histories, fraudulently satisfying the solvency predicate by routing a clean input to an intermediate address and subsequently appending an unrelated, non-compliant transfer originating from that same address. To preclude this, the circuit must enforce explicit Edge Linkage ($\mathsf{receiver}(\tx_{e_{\text{in}}}) = \mathsf{sender}(\tx_{e_{\text{out}}})$). Crucially, because native Ethereum transaction encoding omits the sender's address from the deterministic payload, the source identity can only be resolved via ECDSA signature recovery (\texttt{ecrecover}). Consequently, establishing a tamper-proof native asset DAG dictates that \textit{every} transaction must undergo in-circuit sender extraction to guarantee the cryptographic binding of the flow.

\pparagraph{The Vulnerability of Account-Based Provenance} 
It might seem natural to track compliant history using standard EVM balance mappings. However, account-based ledgers treat all tokens as perfectly interchangeable, acting as a ``fungible wormhole'' where the specific origins of funds are instantly lost. An attacker could receive a clean transfer, spend those specific funds elsewhere, and then reuse their lingering ``clean balance'' status to launder dirty money. Our DAG-based approach solves this by treating provenance as a specific path of \textit{intent} rather than a balance state. To prevent substitution, the sub-DAG must be irreversibly consumed by the terminal execution. As detailed in \Cref{sec:double_spend_history}, this necessitates the use of UTXO-style nullifiers to track the consumption of specific on-chain inflows, even when operating on an account-based blockchain.

\pparagraph{Programmable Routing Policies} In our core $\syss$ framework, we deliberately restrict membership proofs to the topological \emph{origins} of the transaction DAG. This design choice is strictly pragmatic: regulated onboarding points (\eg KYC-compliant exchanges) provide static, highly reliable trust anchors. By anchoring compliance to the source rather than checking every hop against a blocklist, we fundamentally neutralize tainting attacks and eliminate the verifier's burden of synchronizing with dynamic state.

However, our system naturally supports much richer, application-specific constraints. We term this capability \emph{programmable routing policies}. Depending on the specific regulatory context, developers can seamlessly extend the circuit to enforce arbitrary predicates on the intermediate addresses ($a_i$) traversed by the funds. For example, a specialized compliance application could require proof that funds were routed exclusively through authorized institutional liquidity pools, completely avoided specific high-risk jurisdictions, or never interacted with unverified cross-chain bridges. Since the entire provenance graph is already available as a private witness to the prover, enforcing these intermediate routing policies simply requires adding local constraints to the step circuit and incurs zero additional interaction overhead with the base ledger.

\section{Implementation and Performance Evaluation}
\label{sec:implementation}

\subsection{Basic Proof System with Monolithic SNARK}
We implemented an Ethereum-facing linear-path fragment of \syss (\eg linear DAG topology) using \texttt{gnark}~v0.14\footnote{\url{https://github.com/consensys/gnark/}} for R1CS compilation. The circuit utilizes the Groth16 proof over the BN254 curve, ensuring compatibility with Ethereum's EIP-197 pairing precompiles.

\pparagraph{Off-Chain Witness Pipeline}
To generate valid inputs for the circuit, we developed an execution-client--aligned off-chain builder. This pipeline constructs per-block transaction tries and ordered Merkle--Patricia inclusion paths that perfectly align with live consensus client \textsf{transactionsRoot} semantics. Origin-set commitments (\eg the MiMC Merkle tree), padded hop vectors, and graph selection for the compliant sub-path are computed strictly off-chain; the zero-knowledge circuits only enforce algebraic predicates over the resulting byte-level witnesses.

\pparagraph{Circuit Architecture}
Our prototype fuses three core subcircuits in \textsf{PoSoF-Prover}, mapping the formal predicates (\S\ref{sec:formal_predicates}) and Ethereum gadgets (\S\ref{sec:nodal_realization}) into a cross-wired constraint system:
\begin{itemize}
    \item \textbf{\textsf{PathCircuit} (Topology \& Solvency).} Enforces \emph{(i) Source Anchoring} via MiMC membership ($a_0 \in c_{\mathcal{C}}$); \emph{(ii) Edge Linkage and Distinctness} via strict address continuity ($s_0{=}a_0$, $r_i{=}s_{i+1}$, $r_{k-1}{=}a_{\mathrm{dep}}$) and pairwise-distinct private transaction digests; and \emph{(iii) Flow Solvency and terminal execution} via monotone block numbers and bounded, decreasing fractional capacities ($v_i \ge v_{i+1}$, $v_{k-1} \ge V_{\total}$).
    
    \item \textbf{\textsf{HeaderChainCircuit} (Canonicality \& Inclusion).} Validates Merkle--Patricia Trie (MPT) transaction/receipt inclusion and a \texttt{parentHash} chain linking back to a public checkpoint $B_{\text{recent}}$. It RLP-decodes headers to extract exact temporal coordinates $h_i$ and computes legacy EIP-155 sighashes in-circuit for signature binding.
    
    \item \textbf{\textsf{ECDSAHopCircuit} (Authorization).} Executes non-native \texttt{secp256k1} recovery and Keccak address derivation. By equality-gating the recovered $\mathsf{sender}(\tx_i)$ and message hash against the \textsf{PathCircuit} topology and MPT-verified sighashes, it securely binds cryptographic authorization directly to the trie-decoded leaf payload.
\end{itemize}

\begin{table}[t]
\centering
\caption{R1CS constraint counts (BN254, gnark v0.14), per circuit. Default repository build: \textsf{PathCircuit} with MiMC origin-tree depth~$8$ and compile-time maximum hops $L_{\max}{=}8$; \textsf{MPTInclusionCircuit} with $\mathsf{MaxMPTDepth}{=}4$, $\mathsf{MaxNodeBytes}{=}600$; \textsf{HeaderChainCircuit} after absorbing per-hop Merkle--Patricia inclusion, leaf-tx Keccak, and legacy EIP-155 sighash binding (PR4).}
\label{tab:r1cs}

\begin{tabular}{l r}
\toprule
\textbf{Circuit} & \textbf{Constraints} \\
\midrule
\textsf{PathCircuit} (linear path, $L_{\max}{=}8$, MiMC depth~$8$) & 25{,}056 \\
\textsf{MPTInclusionCircuit} ($\mathsf{MaxMPTDepth}{=}4$, $\mathsf{MaxNodeBytes}{=}600$) & 1{,}480{,}413 \\
\textsf{HeaderChainCircuit} ($\mathsf{MaxChainLen}{=}4$, $\mathsf{MaxHeaderBytes}{=}700$; $8\times$ hop bundles) & 24{,}139{,}666 \\
\textsf{ECDSASenderCircuit} (\texttt{secp256k1} verify + Keccak address) & 291{,}302 \\
\textsf{KeccakPreimageCircuit} (demonstrator; not in unified prover path) & 187{,}881 \\
\midrule
\textsf{PoSoF-Prover} ($L_{\max}{=}8$; Path + absorbed header chain + $8\times$ ECDSA hop) & 25{,}681{,}426 \\
\midrule
Sum (five circuits compiled independently) & 26{,}124{,}318 \\
\bottomrule
\end{tabular}
\end{table}

\pparagraph{Monolithic Prover and Public Statement}
To optimize constraint complexity, \textsf{PoSoF-Prover} merges the \textsf{PathCircuit}, the absorbed \textsf{HeaderChainCircuit}, and $L_{\max}$ replicated \textsf{ECDSAHopCircuit} instances into a single compilation unit. This unified R1CS applies cross-component simplifications, yielding fewer total constraints (${\approx}25.7 \times 10^6$ at $L_{\max}{=}8$) than the sum of the independently compiled subcircuits (detailed in \Cref{tab:r1cs}). 

The active path length $k \le L_{\max}$ remains strictly private, enforced via selector-gated constraints over a padded witness. To bind the proof to a one-time use of the compliant flow, the monolithic circuit exposes a public statement that directly mirrors our formal definition, adapted for a linear path:
$x = (c_{\mathcal{C}},\, R_H,\, \tx_{\text{target}},\, V_{\total})$,
where $R_H$ acts as the EVM block header anchor.

\begin{figure*}[htbp]
\centering
\captionsetup[subfigure]{justification=centering}

\begin{subfigure}[b]{0.49\textwidth}
    \centering
    \includegraphics[width=\linewidth]{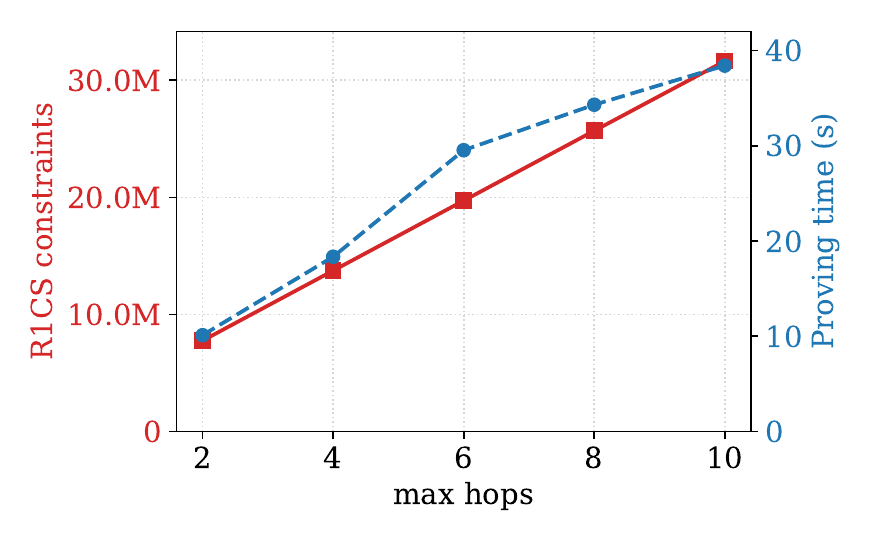}
    \caption{R1CS constraints (left axis) and {proof generation time (right axis).}}
    \label{fig:lmax-r1cs-prove}
\end{subfigure}\hfill
\begin{subfigure}[b]{0.49\textwidth}
    \centering
    \includegraphics[width=\linewidth]{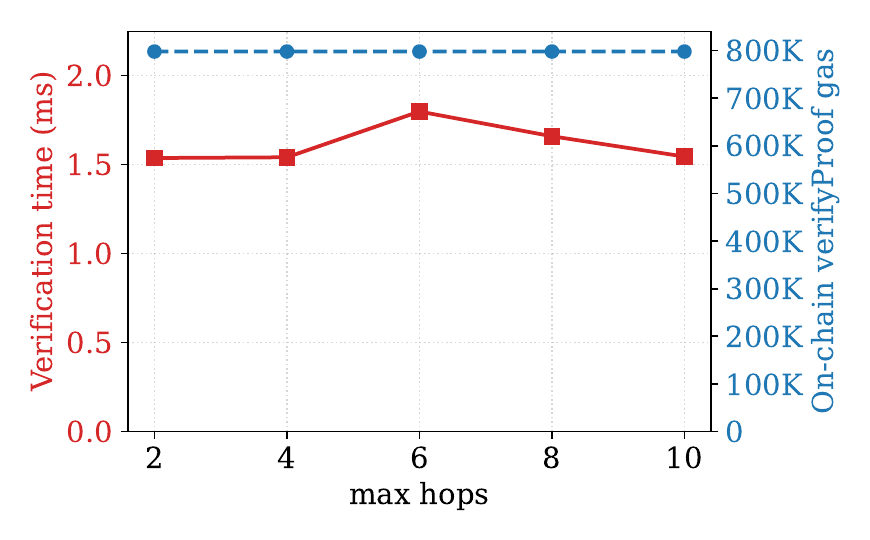}
    \caption{Verification time (left axis) and on-chain \texttt{verifyProof} gas (right axis).}
    \label{fig:lmax-verify-gas}
\end{subfigure}\\[1.5em]


\caption{\textsf{PoSoF-Prover} microbenchmarks vs.\ compile-time maximum hops $L_{\max}$ (active hops $k{=}L_{\max}$ per row; Apple~Silicon laptop; gnark v0.14 Groth16/BN254 harness).}
\label{fig:lmax-sweep}
\end{figure*}

\pparagraph{Performance and EVM Verification} \Cref{tab:bench} and \Cref{fig:lmax-sweep} detail the microbenchmarks over compile-time bounds $L_{\max} \in \{2,4,6,8,10\}$ (active hops $k{=}L_{\max}$ per row). At the default parameterization ($L_{\max}{=}8$, MiMC origin-tree depth $8$), proof generation on an Apple~Silicon laptop takes ${\approx}37$\,s, yielding a compressed $196$\,B proof. 

\begin{table}[t]
\centering
\caption{Groth16 / BN254 microbenchmark for \textsf{PoSoF-Prover} on an Apple~Silicon laptop (single-threaded harness; default $L_{\max}{=}8$, MiMC depth~$8$, $\mathsf{MaxMPTDepth}{=}4$, $\mathsf{MaxChainLen}{=}4$). \textsf{PoSoF-Prover} composes the linear path kernel with the absorbed header chain (per-hop MPT inclusion, leaf-tx Keccak, and legacy EIP-155 sighash binding) and $L_{\max}$ replicated ECDSA hop circuits; the exported verifying key exposes \emph{69} public field elements (path publics plus two $32$-byte digests \texttt{BlockHash} / \texttt{Checkpoint} from the header chain).}
\label{tab:bench}
\begin{tabular}{lr}
\toprule
\textbf{Metric} & \textbf{\textsf{PoSoF-Prover}} \\
\midrule
R1CS constraints                              & 25{,}681{,}426 \\
Setup (compile + trusted setup)               & $\approx 17$\,min \\
Proof generation                              & $\approx 37$\,s \\
Proof verification               & $\approx 1.5$\,ms \\
Proof size (compressed / raw)                 & 196\,B / 388\,B \\
Measured \texttt{verifyProof} gas (local EVM) & $\approx 7.98{\times}10^5$ (${\approx}797{,}916$\,gas)\,\footnotemark[1] \\
\bottomrule
\end{tabular}
\end{table}
\footnotetext[1]{Measured via the same harness (\texttt{gnark} \texttt{vk.ExportSolidity} $\to$ \texttt{solc -{}-optimize -{}-optimize-runs 200} $\to$ \texttt{eth\_estimateGas} on \texttt{go-ethereum}'s \texttt{SimulatedBackend}). The exported \textsf{Verifier} contract exceeds the EIP-170 24\,KiB max code size at \texttt{CREATE}; we install the runtime bytecode at a genesis-allocated address and call \texttt{verifyProof} there. At $L_{\max}{=}8$ the measured gas is $797{,}916$, dominated by the pairing check plus the public-input MSM over $69$ BN254 scalars.}

For on-chain verification, we compiled the exported Solidity verifier via \texttt{solc} (\texttt{--optimize-runs 200}) and profiled it against \texttt{go-ethereum}'s \texttt{SimulatedBackend}. The on-chain \texttt{verifyProof} gas cost is ${\approx~}797{,}916$ at $L_{\max}{=}8$, dominated by the EIP-1108 pairing check and the public-input MSM over $69$ BN254 scalars. We note that at $L_{\max}{=}8$, the resulting bytecode exceeds the EIP-170 $24$\,KiB \texttt{CREATE} limit, necessitating runtime deployment at a genesis-allocated address. While a production deployment scaling to unbounded DAG topologies requires Incrementally Verifiable Computation (IVC) or L2 execution to manage state bloat, this linear-path prototype confirms the EVM-native viability of the core cryptographic predicates.

\subsection{Proof System with Folding-based IVC}
\label{sec:ivc-prototype}

The Groth16 prototype in \S\ref{sec:implementation} compiles the full linear predicate bundle into one monolithic arithmetic circuit. While optimized for a \emph{one-shot} compliance check ($\approx 37$\,s proving time), this static architecture scales poorly for continuous, real-world usage. Because the constraint system is fixed at compile time (allocating $L_{\max}$ hops regardless of actual path length), every new on-chain transfer forces the prover to re-evaluate their entire financial history. 

To achieve proactive, $O(1)$ compliance updates for active wallets, we implemented a parallel proving pipeline utilizing \emph{incrementally verifiable computation (IVC)} via the Nova folding scheme~\cite{kothapalli2022nova}. In this model, the user maintains a folded accumulator; when a new transaction occurs, it is simply folded into the existing proof in $\approx 1.8$\,s. While generating a massive chain strictly from scratch under IVC introduces cumulative cold-start overhead, the $\approx 1.8$\,s marginal update cost represents a definitive leap toward production-grade, continuous on-chain provenance. While our main IVC benchmarks still emphasize linear DAG for comparison with the monolith, we provide an extension to support complex DAG topologies via merge nodes in~\Cref{sec:ivc-dag-impl}.

\pparagraph{The Uniform Step Function ($F$)}
Unlike the monolith, which compiles all hops up to $L_{\max}$ simultaneously, IVC defines a single uniform step $F$. On hop $i$, $F$ consumes an accumulator state $z_i$, a private hop witness $w_i$, and boundary data to output $z_{i+1}$. The folded step uses a fixed arity of five field elements:
$z_i = (\mathsf{holder}_i,\; V_i,\; \mathsf{height}_i,\; \mathsf{tx\_id}_i,\; \mathsf{anchor})$,
where $\mathsf{holder}_i$ is the current address, $V_i$ the compliant value, $\mathsf{height}_i$ the block height, $\mathsf{tx\_id}_i$ a transaction digest, and $\mathsf{anchor}$ a scalar committing to the header context. Each active hop evaluates the exact same R1CS topology, enforcing MPT inclusion, leaf binding, legacy EIP-155 crosslinks, and ECDSA authorization at a cost of $\approx 6.64 \times 10^6$ constraints per step. Crucially, the public statement ($V_{\total}$, $a_{\mathrm{dep}}$, and header digests) and the private witness layout perfectly mirror the monolithic prover defined in \S\ref{sec:implementation}, ensuring identical compliance semantics across both architectures.

\begin{figure*}[t]
\centering
\captionsetup[subfigure]{justification=centering}

\begin{subfigure}[b]{0.49\textwidth}
    \centering
    \includegraphics[width=\linewidth]{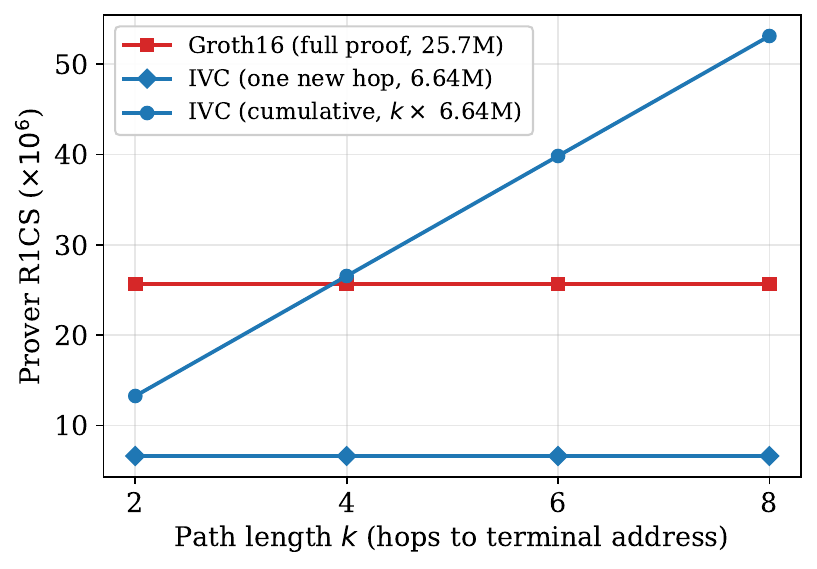}
    \caption{Prover R1CS: Groth16 full proof, IVC one new hop, IVC cumulative ($k$ hops).}
    \label{fig:ivc-r1cs}
\end{subfigure}\hfill
\begin{subfigure}[b]{0.49\textwidth}
    \centering
    \includegraphics[width=\linewidth]{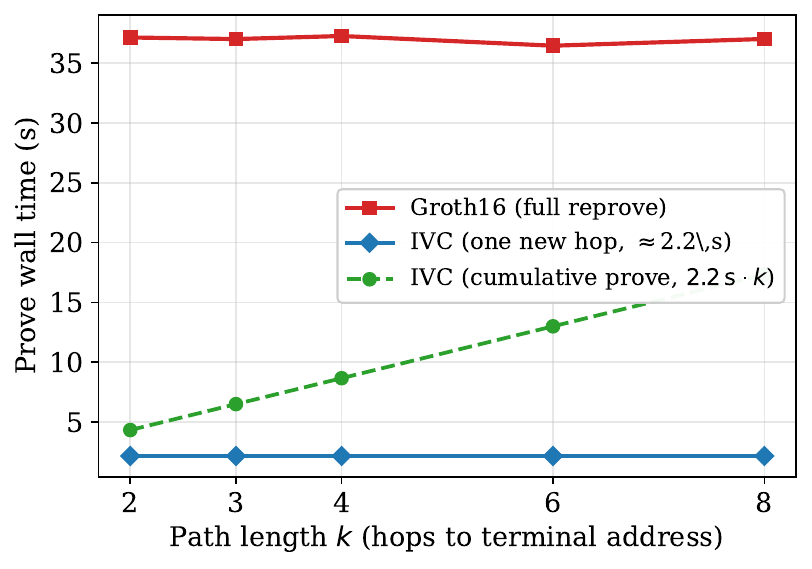}
    \caption{Proving wall time: Groth16 \texttt{Prove}, IVC extend, IVC cumulative (setup excluded).}
    \label{fig:ivc-prove-time}
\end{subfigure}

\caption{Groth16 vs.\ IVC when attesting a provenance path of length $k$ to the terminal address.}
\label{fig:ivc-bench}
\end{figure*}

\pparagraph{Batch vs. Incremental Benchmarks}
\Cref{fig:ivc-bench} illustrates the fundamental trade-off between the two update models. As shown in~\Cref{fig:ivc-r1cs}, the Groth16 monolith allocates a fixed $25.7 \times 10^6$ constraints (padded to $L_{\max}{=}8$); consequently, its proving time remains flat at $\approx 37$\,s regardless of the actual path length $k$ (\Cref{fig:ivc-prove-time}). Conversely, IVC scales linearly. A single hop extension costs $6.64 \times 10^6$ constraints and $\approx 1.8$\,s. If a prover builds a chain entirely from scratch, the cumulative IVC constraints cross the monolithic Groth16 baseline at $k \approx 3.9$ (~\Cref{fig:ivc-r1cs}). Similarly, due to cold-start parameter generation in our Rust prototype, cumulative proving time scales to $120$--$134$\,s. However, as the incremental curve in~\Cref{fig:ivc-prove-time} highlights, this cumulative spike is an artifact of "batching" an entire history at once. For an active, online wallet where folding parameters are cached, the relevant metric is the flat $\approx 1.8$\,s incremental extension, which definitively outperforms the Groth16 monolith for day-to-day state updates.\footnote{Note that Nova (Pallas/Vesta) is not EVM-verifiable, and production deployments wrap the IVC proof in a BN254 Groth16 circuit. Consolidating the public inputs yields an $O(1)$ on-chain verification cost of $\approx 2.4 \times 10^5$\,gas (down from the monolith's $\approx 8.0 \times 10^5$\,gas), independent of DAG complexity}

\section{Related Work}

\label{sec:related}

As regulatory scrutiny on decentralized finance intensifies, a distinct body of literature has emerged around Zero-Knowledge Anti-Money Laundering (ZK-AML)~\cite{piper2025privacy}. These frameworks leverage zero-knowledge proofs to satisfy regulatory requirements---such as proving non-membership in sanctioned address lists or attesting to KYC credentials---without exposing underlying user data to the public ledger. Recently, a concurrent and independent work has identified the need for proactive, prover-side compliance~\cite{duff2026privacy}. However, these proposals largely remain at the architectural level, outlining generalized logic constraints without concrete cryptographic instantiation and performance benchmarking. $\syss$ differentiates itself by bridging this gap: we formalize a concrete cryptographic framework for {transaction provenance} over a generalized value-flow DAG. By coupling this mathematical model with Incrementally Verifiable Computation (IVC), $\syss$ transitions proactive compliance from a theoretical architecture into a highly efficient realization with $O(1)$ verification overhead.

Historically, the tension between privacy and compliance has been most visible in the design of privacy pools, which aim to break the deterministic link between deposit and withdrawal addresses. The notion of a compliant privacy pool was first formalized by Burleson et al.~\cite{burleson2022privacy}. Building on this, Buterin et al.~\cite{buterin2024blockchain} explored the use of zero-knowledge membership and non-membership proofs to construct user-defined ``association sets''~\cite{nadler2023tornado}. This allows users to flexibly prove that their withdrawal originates from a chosen subset of legitimate deposits. However, this retroactive approach is fundamentally vulnerable to \emph{tainting attacks}, wherein an adversary force-transfers illicit funds to honest accounts, systematically polluting their association sets and trapping their funds to obfuscate its own accounts among the actually honest ones. To mitigate this, recent solutions have shifted toward dynamic exclusion lists. Haze~\cite{baranski2023haze} restricts withdrawals for addresses flagged on a dynamically updated blocklist managed via a complaint Merkle tree, while its successor, Daze~\cite{baranski2023haze}, introduces conditional de-anonymization using encrypted nullifiers. Very recently, AuditPay~\cite{tovey2026auditpay} extends blockchain mixers with controlled oversight: rather than requiring users to voluntarily disclose information, it introduces a selective encryption mechanism that cryptographically restricts an auditor's monitoring scope to a committed, hidden budget of addresses per epoch. Unlike Haze and Daze~\cite{baranski2023haze}, which rely on a centralized committee to manage decryption keys, AuditPay enforces its monitoring bound purely through cryptographic assumptions, without introducing additional trust. However, like all post-entry auditing schemes, AuditPay inherits the fundamental limitation that illicit funds may enter the system before auditing begins.

Other notable works focus on tracing or attesting to funds after they have already entered a shielded protocol. Derecho~\cite{beal2024derecho} deploys proof-carrying data~\cite{chiesa2010proof} to propagate allowlist membership proofs through internal pool transfers, enabling institutions to request cryptographic attestations when funds are withdrawn. Similarly, SeDe~\cite{sahu2023sede} proposes a framework to trace and de-anonymize subgraphs of linked illicit transactions within a privacy pool by following the spending of shielded notes. While our DAG traversal conceptually shares similarities with SeDe's subgraph extraction, the architectural philosophy is entirely inverted. Like previous works, Derecho and SeDe also rely on an ``after the event'' retroactive model: illicit funds are permitted to enter the system, and compliance is enforced reactively at the withdrawal phase or via internal de-anonymization. By contrast, $\syss$ acts as a strict admission-control layer. By requiring an efficient proof prior to entry, we prevent illicit funds from ever polluting the target smart contract in the first place, natively neutralizing tainting attacks and eliminating the need for post-facto tracing. We highlight that, $\syss$ and post-deposit mechanisms like Derecho~\cite{beal2024derecho} can be viewed as highly complementary; deploying them in conjunction would provide a comprehensive, end-to-end compliance layer that secures both the entering and exiting of decentralized privacy pools.

\subsection*{Acknowledgments} We are grateful to George Danezis for insightful discussions. István András Seres was supported by the Ministry of Culture and Innovation and the National Research, Development, and Innovation Office within the Quantum Information National Laboratory of Hungary (Grant No. 2022-2.1.1-NL-2022-00004).

\bibliographystyle{plainurl}
\bibliography{references}

\appendix

\section{Zero-Knowledge Argument of Knowledge}\label{sec:zksnarks-formal}

A proof system enables a prover $\prover$ to convince a verifier $\verifier$ about some statement $u$ such that $\exists w: (u; w) \in \rel$, where $w$ is the corresponding witness and $\rel$ is a polynomial-time decidable relation. A proof of knowledge system further convinces the verifier that not only the witness exists, but also the prover knows it. When a proof system only demonstrates some statement holds and does not leak any information about the witness, it is zero-knowledge. A proof system is an argument when (knowledge) soundness only holds against a computationally bounded prover under certain computationally hard assumptions. 

\begin{definition}[Argument System]
 A (non-interactive) argument system $\AS = (\setup, \prove,\break \verify, \simulate)$
 for $\rel$ satisfies the following properties:
 \begin{itemize}
   \item {\bf Completeness.} Given a true statement $u$ for relation $\rel$, an
     honest prover $\prover$ with a valid witness $w$ should convince the verifier
     $\verifier$. More formally, for all $\forall\lambda \in \NN,\forall(u,w)
     \in \rel$:
     $$ \Pr 
       \left[ 
         \AS.\verify(\crs,u,\pi) = 1 ~ \Big | ~
         \begin{aligned}
         (\crs, \td) \leftarrow \AS.\setup(1^\lambda, \rel) \\
         \pi \leftarrow \AS.\prove(\crs, u, w)
         \end{aligned}
       \right] = 1
     $$
   \item {\bf Knowledge Soundness.} There is an extractor that can compute a
     witness whenever the adversary produces a valid argument. The extractor
     gets full access to the adversary's state, including any random coins.
     Formally, we require that $\forall\lambda \in \NN$, $\forall\ \mathsf{PPT}$ adversaries $\adv$ there exists a
     $\mathsf{PPT}$ extractor $\ext_\adv$ such that 
     $$ \Pr 
       \left[ 
       \begin{aligned}
         \AS.\verify(\crs,u,\pi) = 1 \\
         \wedge (u, w) \notin \rel
       \end{aligned}
         ~ \Big | ~
         \begin{aligned}
         (\crs, \td) \leftarrow \AS.\setup(1^\lambda, \rel) \\
           ((u,\pi); w) \leftarrow \adv || \ext_\adv(\crs)
         \end{aligned}
       \right] \le \negl(\lambda)
     $$

   \item {\bf Statistical Zero-knowledge.} An argument is zero-knowledge if it
     does not leak any information besides the truth of the statement.
     Formally, if $\forall\lambda \in \NN,\forall(u, w) \in \rel$ and $\forall\ \mathsf{PPT}$ adversaries $\adv$, the following two distributions are
     statistically close:
     $$
     \begin{aligned}
       D_0 & = \{ \pi_0 \leftarrow \AS.\prove(\crs, u, w): (\crs, \td) \leftarrow \AS.\setup(1^\lambda, \rel) \} \\
       D_1 & = \{ \pi_1 \leftarrow \AS.\mathsf{SimProve}(\crs, \td, u): (\crs, \td) \leftarrow \AS.\mathsf{SimProve}(1^\lambda, \rel)\}
     \end{aligned}
     $$

   \item {\bf Succinctness.} An argument system is said to be \emph{succinct} (\ie SNARK) if there exists a polynomial $p(\cdot)$ such that, for any circuit $C$ of size $|C|$ and any input $u$, the following conditions hold:
$$
T_{\mathsf{Verify}}(C, u, \pi) \le p(\lambda, |u|, \log |C|),
$$
and
$$
|\pi| \le p(\lambda, \log |C|),
$$
where $\lambda$ is the security parameter, $T_{\mathsf{Verify}}$ denotes the verifier's running time, and $|\pi|$ denotes the proof length. Intuitively, this means that both the verification time and the proof size grow \emph{polylogarithmically} in the size of the computation $|C|$ and only \emph{polynomially} in the security parameter and input size.

 \end{itemize}
\end{definition}

\section{Retrospective Local Proving}
\label{sec:instantiation}

Given that the primary targets of our system are public, transparent ledgers (\eg Ethereum and Bitcoin), practical deployment necessitates a \textit{Retrospective Local Proving} architecture. In this model, the generation of the zero-knowledge proof $\pi_{\ori}$ is entirely decoupled from the active participation of intermediate users from source to target. As coordinating with historical senders is practically impossible, the transacting user must act as the sole prover.

To justify the compliant funding of $V_{\total}$ for a target execution payload $\tx_{\text{target}}$, the prover utilizes off-chain graph-search algorithms (\eg backward traversal via archive nodes) to retroactively reconstruct the causal DAG of incoming funds. This extraction terminates when sufficient flow intersects the committed compliant set $\mathcal{C}$. The prover then locally fetches the requisite Merkle state inclusion proofs for all intermediate edges and sequentially executes the zero-knowledge circuits to satisfy the required predicates.

\pparagraph{Advantages and Coordination-Freedom} 
The paramount advantage of this architecture is \textit{immediate deployability}. It requires zero active participation from intermediary senders and imposes no modifications to existing wallet infrastructure. The system leverages the transparency of the base ledger as a data-availability layer for the prover, while utilizing the zero-knowledge property of the resulting proof $\pi_{\ori}$ to completely shield the reconstructed graph topology from the final verifier (\eg the target smart contract).

\pparagraph{Scalability via Trustless Proof Delegation}
A potential downside of the Retrospective Local Proving architecture is the client-side cryptographic burden. While extracting the sub-DAG via constrained graph search is computationally trivial for modern devices, the generation of the zero-knowledge proof for a DAG containing high-frequency transaction edges might be resource-demanding. Specifically, evaluating the circuit constraints for dozens of Merkle-Patricia Trie state inclusions and performing the underlying cryptographic operations (\eg multi-scalar multiplications) might require considerable memory allocation. 

As an alternative approach, our system supports \textit{trustless proof delegation} via a decentralized prover market~\cite{roy2024succinct}. Since the (recursive) SNARK satisfies {knowledge soundness}, a user can securely outsource the heavy computation without compromising security.

\begin{itemize}[leftmargin=*]
    \item \textbf{Delegation:} The user submits a request containing the intended execution payload $\tx_{\text{target}}$ and the target value $V_{\total}$ to a cluster of high-performance provers. 
    \item \textbf{Execution:} The market provers reconstruct the necessary sub-DAG, perform the intensive folding and non-native arithmetic computations, and return the constant-size proof $\pi_{\ori}$.
    \item \textbf{Trustless Verification:} The soundness property guarantees that the delegated prover cannot forge a false provenance certificate. Also, recent private delegation techniques~\cite{abbaszadeh2025single, chiesa2023eos} allow for preserving the privacy of the user by hiding the witness (\ie sub-DAG) and even the proof itself. Due to {terminal execution binding}, the proof is only valid for the user's specific $\tx_{\text{target}}$. The user simply verifies $\pi_{\ori}$ locally in constant time (\eg milliseconds) before broadcasting their final transaction to the ledger.
    
\end{itemize}

This delegation model effectively shifts the computational complexity of compliance from the end-user to specialized infrastructure, drastically improving the accessibility and user experience of privacy-preserving compliance on transparent ledgers.

\section{DAG Reconstruction Heuristics}
To instantiate our architecture, the client must extract a valid compliant sub-DAG $\mathcal{G}' \subseteq \mathcal{G}$ from their historical on-chain activity. While the prover naturally possesses off-chain knowledge of their own transaction history, isolating the \emph{optimal} subset of transactions---specifically, the sub-DAG with the minimum number of edges that satisfies the target flow solvency---is of importance for minimizing the constraint of the resulting \syss proof. 

As discussed before, while resolving dense, highly co-mingled transaction topologies is formally equivalent to a maximum network flow problem, the vast majority of user transaction histories are sparse. For these standard cases, we adapt a standard Breadth-First Search (BFS) into a capacity-prioritized heuristic, detailed in Algorithm~\ref{alg:retro_search}.\footnote{We remark that compliant subgraph extraction extends beyond folklore path-finding algorithms (\eg Dijkstra or Algorithm~\ref{alg:retro_search}). Because ledger transactions are highly co-mingled and subject to fractional capacity bounds, isolating a single shortest path is often insufficient. The prover must execute a constrained search---functionally equivalent to a bounded network flow problem---to extract a multi-source sub-DAG that collectively satisfies Flow Solvency.} The algorithm utilizes an archive node to traverse edges in reverse chronological order. By filtering out micro-transactions (transfers below a threshold $\tau$) and prioritizing incoming edges with the highest compliant value, the heuristic actively minimizes the in-degree of every resolved node. Crucially, it maintains a global tracker of requested capacity per edge to natively prevent over-drafting when multiple branches intersect at a shared intermediate transaction. This optimization directly reduces the size of $\mathcal{G}'$, thereby minimizing the computational overhead of proof generation.

\begin{algorithm}[t]
\small
\caption{Capacity-Prioritized Provenance Search (Fast-Path Heuristic)}
\label{alg:retro_search}
\KwIn{Target execution $\tx_{\text{target}}$, value $V_{\total}$, compliant set $\mathcal{C}$, threshold $\tau$}
\KwOut{Provenance sub-DAG $\mathcal{G}' = (\mathcal{V}', \mathcal{E}')$ or $\bot$}

$n_{\text{target}} \gets (\mathsf{sender}(\tx_{\text{target}}), \mathsf{seq}(\tx_{\text{target}}))$\;
$\mathcal{V}' \gets \{n_{\text{target}}\}, \ \mathcal{E}' \gets \emptyset, \ Q \gets [(n_{\text{target}}, V_{\total})]$\;
$\mathsf{GlobalReq} \gets \text{Map}()$ \tcc*{Tracks capacity bounds for overlapping paths}

\While{$Q \neq \emptyset$}{
    $((a, h), v_{\text{req}}) \gets Q.\textsf{Dequeue}()$\;
    
    \If{$a \in \mathcal{C}$}{
        \textbf{continue}\;
    }
    
    $\mathcal{E}_{\text{in}} \gets \textsf{SortDesc}(\{e \in \textsf{Query}(a, h) \mid \mathsf{val}(e) \ge \tau \}, \mathsf{val})$\;
    
    \For{$e = (n_{\text{src}}, (a,h), v) \in \mathcal{E}_{\text{in}}$}{
        \tcc{Calculate available fractional capacity to prevent edge overdraft}
        $v_{\text{avail}} \gets v - \mathsf{GlobalReq}.\textsf{Get}(e, 0)$\;
        \If{$v_{\text{avail}} \le 0$}{ \textbf{continue}\; }
        
        $\mathcal{V}' \gets \mathcal{V}' \cup \{n_{\text{src}}\}, \ \mathcal{E}' \gets \mathcal{E}' \cup \{e\}$\;
        $v_{\text{alloc}} \gets \min(v_{\text{avail}}, v_{\text{req}})$\;
        
        $\mathsf{GlobalReq}.\textsf{Add}(e, v_{\text{alloc}})$\; 
        $Q.\textsf{Enqueue}((n_{\text{src}}, v_{\text{alloc}}))$\;
        $v_{\text{req}} \gets v_{\text{req}} - v_{\text{alloc}}$\;
        
        \If{$v_{\text{req}} \le 0$}{
            \textbf{break}\;
        }
    }
    
    \If{$v_{\text{req}} > 0$}{
        \Return $\bot$ \tcc*{Heuristic failed, fallback to Max-Flow required}
    }
}

\Return $(\mathcal{V}', \mathcal{E}')$\;
\end{algorithm}
\footnotetext{Crucially, the prover retains complete autonomy in selecting which valid edges to include in the witness. The zero-knowledge circuit enforces verification, not determinism; as long as the extracted topology satisfies the predicates (\eg Temporal Ordering, Flow Solvency) and successfully routes $V_{\total}$ to the target, the proof is valid. This flexibility allows provers to optimize for computational efficiency or actively obscure specific funding routes.}

\section{Application-Layer Defenses: Double-Spending of Clean History}
\label{sec:double_spend_history}

While the solvency predicate natively prevents the arbitrary creation of compliant funds within a single, isolated DAG,  proofs are inherently stateless. A computationally~valid proof $\pi_{\ori}$ merely asserts a historical truth; it does not possess intrinsic awareness of its own prior use. This introduces a critical system-level threat: {Double-Spending of Clean History}.

\pparagraph{Attack Scenario}
Suppose a user, Alice, reconstructs a valid provenance DAG demonstrating the receipt of $100$ compliant ETH from a compliant source. Alice successfully generates a proof $\pi_1$ and deposits $100$ ETH into Privacy Pool A. Because her on-chain transaction history remains cryptographically valid indefinitely, Alice could subsequently generate a second proof $\pi_2$ (utilizing the exact same historical sub-DAG as the private witness) to deposit another $100$ ETH into Privacy Pool B. Despite possessing only $100$ ETH of genuinely compliant origin capacity, Alice has successfully laundered $200$ ETH.

\pparagraph{Cryptographic Nullifiers and Change Commitments}
To mathematically preclude both Capacity Amplification and Dirty Substitution, the provenance DAG cannot terminate arbitrarily at the user's transparent address. The final evaluated edge must cryptographically bind to the specific execution of the privacy pool deposit. Furthermore, to prevent the reuse of the origin sub-DAG across multiple deposits without compromising the zero-knowledge property, the verifying smart contract must enforce state transitions utilizing a UTXO-style internal architecture, regardless of whether the underlying base-ledger is account-based (\eg Ethereum) or UTXO-based.

For applications requiring strict consumption of historical flows, the $\syss$ circuit is parameterized to output deterministic \textit{Nullifiers} as public instance variables. Crucially, the nullifier must uniquely identify the specific compliant inflows being consumed, rather than the origin source itself, to prevent collisions. Let $\mathcal{E}_{\text{in}}(a)$ be the set of edges terminating at the prover's address $a$. To prevent a \emph{partial replay attack}---where an adversary mixes previously consumed edges with new ones to generate a novel, monolithic nullifier---the circuit must output a \emph{set} of individual nullifiers, one for each consumed terminal edge: $\mathsf{Null}_e = \mathsf{Hash}(\tx_e, \mathsf{sk})$ for all $e \in \mathcal{E}_{\text{in}}(a)$. Upon successful proof verification, the smart contract records these nullifiers in a public registry, ensuring any subsequent proof attempting to claim even a single previously used inflow will revert.

This binary consumption, however, introduces a \emph{Capacity Destruction} problem: consuming a large historical inflow for a small target deposit would permanently destroy the remaining clean history. For applications requiring highly flexible, fractional deposits, the circuit resolves this internally via \textit{private change commitments}. If a user deposits $V_{\text{dep}}$ from a proven historical capacity of $V_{\text{total}}$ (where $V_{\text{dep}} < V_{\text{total}}$), the circuit generates the public nullifiers to irrevocably consume the underlying inflows, alongside a new, private change commitment for the remainder ($V_{\text{total}} - V_{\text{dep}}$). The smart contract appends this change commitment to an on-chain Merkle tree. For future transactions, the user anchors their proof directly to this private change commitment rather than reconstructing a new path from the public ledger.

\begin{remark}
    One might intuitively attempt to secure a decoupled capacity state by incorporating temporal constraints (\eg tracking the block height of the clean inflow). However, this fails to prevent \textit{dirty substitution} due to the fungible nature of EVM account balances. Because smart contracts cannot natively monitor continuous historical balances without prohibitive gas overhead, an adversary can attest to a clean inflow at $t_1$, empty their account at $t_2$, and launder dirty funds at $t_3$. To achieve security, the provenance sub-DAG must atomically terminate at the execution of the deposit transaction itself, necessitating the use of UTXO-style nullifiers and change commitments to manage the resulting fractional state.
\end{remark}

\pparagraph{Target-Source Unlinkability}
While the mechanisms above successfully mitigate the double-spending of historical capacity, the introduction of deterministic public variables---such as nullifiers and on-chain change commitments---necessitates strict security definitions that extend beyond the core $\syss$ protocol. As established in literature, such as \textit{Ledger Indistinguishability} in Zerocash~\cite{sasson2014zerocash} and \textit{Unlinkability} in anonymous credentials~\cite{camenisch2001efficient}, perfect zero-knowledge at the proof level does not inherently guarantee protocol-level privacy. If application-layer variables are improperly randomized, an adversary could trivially link transactions despite the underlying proof revealing nothing.

To formalize this protection, our protocol enforces \textit{Target-Source Unlinkability}. This property guarantees that an adversary observing multiple public target executions ($\tx_{\text{target}}$) and their associated state transitions cannot link them to their specific origin sources, nor determine if they share overlapping intermediate topologies (\ie overlapping sub-DAGs, shared intermediary nodes). We can define this via a standard indistinguishability game: an adversary submits two valid, compliant sub-DAGs ($w_0, w_1$) that both satisfy the capacity constraints for a target execution. The challenger selects a random bit $b \in \{0, 1\}$, generates the zero-knowledge proof and corresponding application-layer state (nullifiers/commitments) using $w_b$, and returns them. The system satisfies Target-Source Unlinkability if the adversary's advantage in guessing $b$ is strictly negligible ($\le \mathsf{negl}(\lambda)$). By deriving our private change commitments through collision-resistant hashing with user-specific randomness, our architecture natively satisfies this protocol-level guarantee.

\begin{table}[!t]
    \centering
    \small 
    \caption{Threat Model and Predicate Mapping by Security Dimension}
    \label{tab:threat_model}
    \begin{tabular}{@{} c >{\raggedright\arraybackslash}p{4.2cm} >{\raggedright\arraybackslash}p{3.4cm} @{}}
        \toprule
        & \textbf{Attack Vector} & \textbf{Mitigating Predicate} \\
        \midrule

        \multirow{10}{*}{\rotatebox[origin=c]{90}{\textbf{Existence}}} &
        \textbf{Ledger Fabrication} \newline 
        \textit{\scriptsize Fabricating transactions that do not exist on the canonical chain.} & 
        \textbf{On-chain Integrity} \newline 
        Inclusion Proof ($\Pi$) \\ \addlinespace[0.15cm]
        
        & \textbf{Capacity Forgery} \newline 
        \textit{\scriptsize Claiming compliant volume exceeding actual ledger records.} & 
        \textbf{Fractional Bounds} \newline 
        $v_e \le v(\tx)$ \\ \addlinespace[0.15cm]
        
        & \textbf{Malicious Origin Injection} \newline 
        \textit{\scriptsize Laundering unauthorized funds into the proven sub-DAG.} & 
        \textbf{Source Anchoring} \newline 
        $v_{0} \in \mathcal{C}$ \\ \midrule
        
        \multirow{10}{*}{\rotatebox[origin=c]{90}{\textbf{Continuity}}} &
        \textbf{Topology Forgery} \newline 
        \textit{\scriptsize Stitching disconnected transactions to fake a path.} & 
        \textbf{Edge Linkage} \newline 
        $\mathsf{recv}(e_{\text{in}}) = \mathsf{send}(e_{\text{out}})$ \\ \addlinespace[0.15cm]

        & \textbf{Chronological Manipulation} \newline 
        \textit{\scriptsize Time-travel exploits to use future funds to justify past actions.} & 
        \textbf{Temporal Ordering} \newline 
        $h(e_{\text{in}}) < h(e_{\text{out}})$ \\ \addlinespace[0.15cm]

        & \textbf{Flow Amplification} \newline 
        \textit{\scriptsize Creating value mid-routing by outputting more than received.} & 
        \textbf{Flow Solvency} \newline 
        $\sum v_{\text{out}} \le \sum v_{\text{in}}$ \\ \midrule
        
        \multirow{9}{*}{\rotatebox[origin=c]{90}{\textbf{Non-Reusability}}} &
        \textbf{Intra-DAG Path Overlap} \newline 
        \textit{\scriptsize Recycling a transaction across branches within the same proof.} & 
        \textbf{Distinctness} \newline 
        $\tx_i \neq \tx_j$ \\ \addlinespace[0.15cm]
        
        & \textbf{Double-Spending \& Replay} \newline 
        \textit{\scriptsize Reusing a historical path in a new proof, or hijacking a payload.} & 
        \textbf{Terminal Execution \& Nullifiers} \newline 
        Cryptographically bound to $\tx_{\text{target}}$ \\
        
        \bottomrule
    \end{tabular}
\end{table}

\section{\sys Instantiation for Bitcoin}\label{sec:bitcoin_instantiation}
In this section, we briefly outline the architectural modifications required to instantiate \syss for Bitcoin~\cite{nakamoto2008bitcoin}. While the core philosophy of prover-side provenance remains identical, Bitcoin’s Unspent Transaction Output (UTXO) model fundamentally alters the topological mapping of the provenance graph, simplifying certain predicates while introducing new cryptographic complexities for the step circuit.

\begin{description}[leftmargin=*]
    \item[Native UTXO Graph Topology and Flow Solvency] 
    Unlike the Ethereum account-based model---where graph nodes represent addresses and edges represent transactions---the Bitcoin ledger natively forms a strict Directed Acyclic Graph. In a UTXO-based \syss instantiation, the nodes in the provenance sub-DAG map directly to \emph{transactions}, and the directed edges represent discrete \emph{UTXOs}. Because a UTXO can only be consumed in its entirety, the flow solvency predicate at any transaction node simply requires proving that the sum of created outputs is less than or equal to the sum of consumed inputs (accounting for the implicit miner fee): $\sum_{e_{\text{out}} \in \mathcal{E}_{\text{out}}(n)} v_{e_{\text{out}}} \le \sum_{e_{\text{in}} \in \mathcal{E}_{\text{in}}(n)} v_{e_{\text{in}}}$. This structural rigidity natively enforces transaction distinctness and eliminates the ambiguity of account-balance mixing. \medskip

    \item[$\mathsf{VerifyChainToHead}(B, H)$ via NIPoPoWs] 
    After verifying that a specific transaction exists in a block $B$ via a standard Merkle inclusion proof ($\mathsf{VerifyTxInclusion}$), the prover must establish that $B$ belongs to the canonical chain. Rather than traversing every block header linearly, this can be proven succinctly using Non-Interactive Proofs of Proof-of-Work (NIPoPoWs)~\cite{kiayias2020non}. Recent advancements~\cite{aumayr2025blink} enable $\mathcal{O}(1)$-sized NIPoPoWs, allowing the \syss circuit to verify canonical chain inclusion with a constant number of constraints, entirely independent of the blockchain's total length or the depth of the historical block $B$.\medskip

    \item[Cryptographic Heterogeneity] 
    Bitcoin currently supports multiple address standards (Legacy, native/nested SegWit, and Taproot) and two distinct digital signature schemes (ECDSA and Schnorr). Consequently, the uniform step circuit in \syss must be equipped to parse these varying transaction witness formats dynamically. Specifically, the circuit must conditionally route to either secp256k1 ECDSA or Schnorr signature verification modules. This imposes non-trivial circuit complexity and constraint overhead due to the extensive use of non-native field arithmetic required to evaluate these signatures inside the SNARK curve.
\end{description}

\pparagraph{Verification Architecture}
Finally, it is crucial to note that Bitcoin Script currently lacks the expressivity required to natively verify succinct proofs (\eg Groth16) on-chain. Consequently, a Bitcoin-native \sys implementation necessitates an asymmetric verification architecture. We assume the ultimate verifier is either an off-chain compliance monitor (\eg a regulated exchange enforcing admission control prior to custodial deposits) or an EVM-compatible cross-chain bridge. In this paradigm, the Bitcoin ledger serves purely as the immutable data availability and provenance layer, while the $\mathcal{O}(1)$ cryptographic proof is evaluated externally.

We leave the concrete implementation, constraint optimization, and performance benchmarking of a UTXO-native \syss prover as an exciting direction for future work.

\section{Extended Evaluation: IVC-based DAG Realization}
\label{sec:ivc-dag-impl}

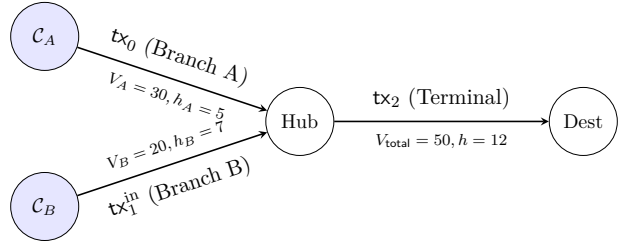
\begin{figure}[t]
    \centering
    \scalebox{0.75}{
    \begin{tikzpicture}[
        >=stealth,
        vertex/.style={circle, draw, minimum size=1.2cm, inner sep=2pt, font=\small},
        edge/.style={->, thick}
    ]
    \node[vertex, fill=blue!10] (origin1) at (0, 1.5) {$\mathcal{C}_A$};
    \node[vertex, fill=blue!10] (origin2) at (0, -1.5) {$\mathcal{C}_B$};
    \node[vertex] (hub) at (4.5, 0) {Hub};
    \node[vertex] (deposit) at (9.5, 0) {Dest};

    \draw[edge] (origin1) -- node[above=2pt, sloped] {$\mathsf{tx}_0$ (Branch A)} 
                             node[below=2pt, sloped] {\scriptsize $V_A=30, h_A=5$} (hub);
                             
    \draw[edge] (origin2) -- node[below=2pt, sloped] {$\mathsf{tx}_1^{\text{in}}$ (Branch B)} 
                             node[above=2pt, sloped] {\scriptsize $V_B=20, h_B=7$} (hub);
                             
    \draw[edge] (hub) -- node[above=2pt] {$\mathsf{tx}_2$ (Terminal)} 
                         node[below=2pt] {\scriptsize $V_{\total}=50, h=12$} (deposit);
    \end{tikzpicture}
    }
    \caption{Toy \texttt{dag\_diamond} topology. Two compliant origins ($\mathcal{C}_A, \mathcal{C}_B$) converge at an intermediate merge hub. Rather than utilizing dual independent proofs, our circuit multiplexing strategy folds this convergence into a single incrementally verifiable chain.}
    \label{fig:dag_diamond}
\end{figure}

The prototype evaluated in \S\ref{sec:implementation} demonstrates the core $\syss$ protocol over a linear sequence of transaction hops. However, real-world provenance histories frequently manifest as proper directed acyclic graphs (DAGs), where multiple compliant branches of funds converge at intermediate addresses prior to a terminal deposit. While fully generalized non-linear topologies typically require heavy Proof-Carrying Data (PCD) or tree-folding frameworks, this appendix details how we extended our Nova-based stack to natively support bounded DAG convergences (\eg diamond patterns as in~\Cref{fig:dag_diamond}) \emph{inside a single} incrementally verifiable chain.

\subsection{Linearizing the DAG via Witness Serialization}
A na\"ive approach to graph traversal would force the prover to maintain two separate SNARK chains up to the merge hub, requiring an expensive terminal proof verification for each. To circumvent this, our prover engine performs a topological sort of the sub-DAG, flattening the execution into an ordered list of linear steps. 

When a convergence occurs, we employ a witness serialization trick. The primary Nova accumulator persistently tracks the state of the dominant path (Branch A). When the traversal reaches the merge hub, the prover supplies the verified state of the concurrent path (Branch B) entirely as a private witness array to the step circuit. This allows the circuit to mathematically absorb the incoming liquidity without breaking the linear chain of the folding scheme.

\subsection{Circuit Multiplexing}
Because Nova strictly requires a uniform step function $F$ for every fold, we multiplexed the step logic. The universal $\syss$ step circuit dynamically toggles between two constraint families based on a private step-kind flag:
\begin{itemize}[leftmargin=*]
    \item $F_{\mathsf{hop}}$ (\textbf{Standard Edge}): Evaluates a standard on-chain transaction (\eg $\mathsf{tx}_0$ or $\mathsf{tx}_2$), executing the full cryptographic suite described in \S\ref{sec:implementation} (MPT inclusion proofs, Keccak state digests, and secp256k1 recovery).
    \item $F_{\mathsf{merge}}$ (\textbf{Merge Hub}): Pauses the heavy on-chain cryptographic checks. Instead, it enforces the nodal \texttt{merge\_collect} predicates. It verifies that the accumulator state (Branch A: $V_A, h_A$) and the witnessed state (Branch B: $V_B, h_B$) share the identical intermediate target address. It then outputs the merged value $V_{\total} = V_A + V_B$ and the maximal temporal bound $\max(h_A, h_B)$ to preserve chronological ordering.
\end{itemize}
By gating the heavy constraints during $F_{\mathsf{merge}}$, the circuit remains logically uniform but computationally asymmetric, avoiding unnecessary EVM state verifications when merely combining pre-verified branches.

\subsection{Performance and Microbenchmarks}
\Cref{tab:ivc-dag-bench} reports the end-to-end \texttt{prove + verify} times for topological graph routing versus full cryptographic enforcement. The \emph{Graph} mode strictly isolates the step-multiplexing and topological constraints (\eg \texttt{merge\_collect}), eliminating EVM primitives to isolate the structural cost. The \emph{Crypto} mode runs the full Ethereum verification suite over the multiplexed paths.

\begin{table}[ht]
    \centering
    \caption{IVC DAG prototype: median cold-batch \texttt{prove+verify} time (seconds) on Apple Silicon. {Crypto} times are dominated by the one-time recursive CRS setup ($\sim 100$ seconds per cold run), independent of hop count.}
    \label{tab:ivc-dag-bench}
    \small
    \begin{tabular}{@{}lp{4.8cm}rrr@{}}
        \toprule
        \textbf{Workload} & \textbf{Proven path (summary)} & \textbf{Steps} & \textbf{Graph(s)} & \textbf{Crypto(s)} \\
        \midrule
        \texttt{dag\_diamond} & $\mathsf{tx}_0$, merge$(\mathsf{tx}_1^{\mathrm{in}})$, $\mathsf{tx}_2$ & 3 & 1.10 & --- \\
        \texttt{dag\_diamond\_crypto} & $\mathsf{tx}_0$, merge, $\mathsf{tx}_2$, $\mathsf{tx}_3$ & 4 & --- & 127.8 \\
        \midrule
        \texttt{linear\_k3} & $\mathsf{tx}_0, \mathsf{tx}_2, \mathsf{tx}_3$ (no merge) & 3 & 1.11 & 130.8 \\
        \texttt{linear\_k8} & Eight consecutive hops & 8 & 1.26 & 136.2 \\
        \bottomrule
    \end{tabular}
\end{table}

The absolute times reported in the Crypto column ($\approx 128$--$136$\,s) represent a "cold-start" batch, which is heavily dominated by the one-time folding parameter setup, taking approximately 100 seconds. Once parameters are amortized (as expected in a real-world local wallet instance), the marginal cost of proving an additional hop remains $\approx 2.2$ seconds, similar to our linear benchmarks.

Crucially, the data demonstrates the efficiency of the DAG multiplexing approach. Proving the realistic Ethereum diamond (\texttt{dag\_diamond\_crypto}) takes slightly \emph{less} time than the pure 3-hop linear equivalent (\texttt{linear\_k3}), despite having an extra folding step. This is because the $F_{\mathsf{merge}}$ row successfully gates off the Keccak/MPT constraints, allowing the prover to logically converge funds without incurring the heavy cryptographic penalty of processing redundant state-roots.

\end{document}